\documentclass[useAMS,usenatbib]{mnras}
\usepackage{graphicx}
\usepackage{amsmath}
\usepackage{amssymb}
\usepackage{color}
\usepackage[T1]{fontenc}
\usepackage{times}

\def \msun {\ifmmode \rm M_{\odot} \else $\rm M_{\odot}$ \fi}

\title[The hidden discs of simulated galaxies]{NIHAO VI. The hidden discs of simulated galaxies}
 \author[Obreja et al]{Aura Obreja$^{1,2}$\thanks{E-mail: obreja@mpia.de}, Gregory S. Stinson$^{2}$, Aaron A. Dutton$^{1,2}$, Andrea V. Macci\`{o}$^{1,2}$, \and Liang Wang$^{2,3,4}$ and Xi Kang$^{3,4}$\\  
 $^{1}$New York University Abu Dhabi, PO Box 129188, Saadiyat Island, Abu Dhabi, United Arab Emirates\\
 $^{2}$Max Planck Institute f\"{u}r Astronomie, K\"{o}nigstuhl 17, 69117 Heidelberg, Germany\\
 $^{3}$Purple Mountain Observatory, the Partner Group of MPI f\"{u}r Astronomie, 2 West Beijing Road, Nanjing 210008, China\\
 $^{4}$University of Chinese Academy of Science, 19A Yuquan Road, 100049 Beijing, China}

\begin{document}

\maketitle

\label{firstpage}

\begin{abstract}
Detailed studies of galaxy formation require clear definitions of the structural components of galaxies.
Precisely defined components also enable better comparisons between observations and simulations. 
We use a subsample of eighteen cosmological zoom-in simulations  from the {\sc NIHAO} project
to derive a robust method for defining stellar kinematic discs in galaxies.
Our method uses Gaussian Mixture Models in a 3D space of dynamical variables. 
The {\sc NIHAO} galaxies have the right stellar mass for their halo mass, and their angular momenta and S\'{e}rsic indices match observations. 
While the photometric disc-to-total ratios are close to 1 for all the simulated galaxies, the kinematic ratios are around $\sim$0.5.
Thus, exponential structure does not imply a cold kinematic disc. 
Above M$_{\rm *}\sim10^{\rm 9.5} \msun$, the decomposition leads to thin discs and spheroids that have clearly different properties, 
in terms of angular momentum, rotational support, ellipticity, [Fe/H] and [O/Fe].   
At M$_{\rm *}\lesssim10^{\rm 9.5} \msun$, the decomposition selects discs and spheroids with less distinct properties.
At these low masses, both the discs and spheroids have exponential profiles with high minor-to-major axes ratios, i.e. thickened discs.  

\end{abstract}


\section{Introduction}

Galaxies are comprised of several components, including structures such as the thin and thick discs, the bulge, and the stellar halo. 
Understanding how galaxies form and evolve requires understanding what processes create each component
and how material moves from one component to another. 
Studying each component separately requires a clear definition for each.

Broadly speaking the components of galaxies are classified into two larger categories, discs and spheroids.
Five main properties distinguish discs from spheroids:
\begin{itemize}
\item {\bf Shape:}  Discs are flattened with one axis much shorter than the other two perpendicular axes that are of about the same length. 
Spheroids are round with three perpendicular axes that are all about the same length.
\item {\bf Kinematics:}  Discs are rotationally supported.  
The dominant component of disc stellar velocities is in the azimuthal direction.  
Spheroids are dispersion supported such that stars follow random orbits.
\item {\bf Age:}  Discs often consist of young stars giving them blue colors. 
Spheroids stopped forming stars Gyrs ago such that they have red colors.
\item {\bf Metal abundance:}  Metal abundance closely follows stellar ages such that young discs have lower relative 
$\alpha$-enrichment than spheroids.
\item {\bf Surface density profile:}  Discs have mass surface density distributions well described by exponentials,
while spheroids follow de Vaucouleurs profiles. 
In typical galaxies, exponential profiles dominate the outskirts of galaxies while steeper de Vaucouleurs
spheroids dominate the centres.
\end{itemize}

Before making detailed studies of the origin of galactic structures, it is necessary to identify the individual components. 
For a long time, galaxies were decomposed based only on their photometric appearance 
\citep{deVaucouleurs:1959, vanHouten:1961, Freeman:1970, Yoshizawa:1975, Simien:1989}.  
Disc shaped galaxies with regular rotation curves mostly exhibit exponential surface density profiles. 
So, galaxies with exponential density profiles are usually assumed to have stars on regular circular orbits.

Discs and spheroids can be classified more robustly using their dynamical properties.
Discs are rotation supported, while spheroids are dispersion supported.
Precise stellar kinematics, however, can only be measured for the Milky Way. 
In external galaxies, the maximum dynamical information is provided by Integral Field Unit (IFU) spectroscopy 
in the form of 2D line-of-sight velocity moments maps
\citep[for example][]{Bacon:2001, Bershady:2010, Cappellari:2011, Sanchez:2012, Bundy:2015}.
Each spectral element (spaxel) contains the combined light of stars on a variety of orbits. 
To determine the relative proportions of stars in each galaxy component, 
one can construct equilibrium model galaxies out of stars selected from a variety of families and energies 
\citep[a technique called ``Schwarzschild modeling'',][]{Schwarzschild:1979}. 
For a recent example of Schwarzschild modeling using IFU data, see \citet{vandenBosch:2009}.
As is an issue with many observations, young stars are much brighter than old ones, so when a galaxy contains a young stellar population, 
it makes it difficult to detect the kinematics of the old population. 
It is thus likely that the disc component of galaxies containing young stellar populations will be emphasized.

There has been significant theoretical work on how stellar discs and spheroids form in the context of galaxy formation.
As neighboring haloes collapse, their tidal torques provide the gas with angular momentum \citep{Peebles:1969, White:1984}.
Initially, gas and dark matter have a common angular momentum.
In the standard picture, stellar discs form from gas that cools and settles into centrifugal equilibrium 
at the bottom of the potential energy within a spinning dark matter halo \citep{White:1978, Fall:1980, Efstathiou:1983}. 
Since the gas is collisional it stays on circular orbits, thus the stars also form with circular orbits. 
Discs form in such models with exponential profiles over a few scalelengths \citep{Dalcanton:1997, Dutton:2009}.
These disc galaxies follow the observed Tully-Fisher relation and result 
in a scatter on the size-rotation velocity plane consistent with observations \citep{Mo:1998}.

Conventionally, discs increase their vertical velocity dispersion through processes like minor mergers 
\citep{Quinn:1993, Walker:1996, Velazquez:1999, Berentzen:2003, Read:2008, Villalobos:2008, Kazantzidis:2009, Moster:2010},
radial migration \citep{Loebman:2011, Schonrich:2009, Scannapieco:2011}
or scattering by dense mass clumps \citep{Spitzer:1953, Lacey:1984}.
In this way, old stellar populations comprise a thick disc and younger generations constantly reform a thin disc.
Recent numerical simulations indicate that the picture might not be so straightforward. Discs might form thick 
out of a turbulent formation environment caused by merging \citep{Brook:2004}, clumpy disc structure \citep{Bournaud:2009}, 
or stellar feedback \citep{Bird:2013, Stinson:2013b}.

In a hierarchical picture of galaxy formation, 
mergers could disrupt thin discs so extensively as to produce the spheroid component \citep{Davies:1983, Hopkins:2009, Brooks:2015, Kannan:2015}.
The prevalence of thin discs in the Universe, though, implies that the effect of mergers is limited or that 
mergers could help reform discs \citep[e.g. gas rich mergers,][]{Robertson:2006, Hammer:2009}.
Post-merger smooth gas infall could also reform thin discs \citep[e.g.][]{Villalobos:2010, Puech:2012, Moster:2012}.

Numerical simulations of galaxy formation have improved considerably in the last decade
\citep{Brook:2012, Aumer:2013, Stinson:2013a, Christensen:2014, Hopkins:2014, Kannan:2014, Marinacci:2014, Schaye:2015, Wang:2015}.
The important driver of these advances was the implementation of effective stellar feedback 
\citep{Stinson:2006, DallaVecchia:2008}. 
Stellar feedback can prevent the catastrophic loss of angular momentum that plagued earlier simulations \citep[e.g.][]{Navarro:2000}. 
In this manner, simulations now reproduce many of the statistical properties of galaxy populations 
\citep{Aumer:2013, Hopkins:2014, Vogelsberger:2014, Schaye:2015, Wang:2015}.

In particular, the large sample of cosmological ``zoom-in'' simulations presented by \citet{Wang:2015}
reproduces the stellar mass - halo mass relation \citep{Behroozi:2013, Moster:2013, Kravtsov:2014} over four orders of magnitude at various redshifts. 
The feedback model used \citep{Stinson:2006, Stinson:2013a} 
is able to reproduce a wide range of observational constrains \citep{Brook:2012, Stinson:2013a, Stinson:2013b}, 
not only for Milky Way mass systems but also from dwarfs up to systems a few times more massive than the Milky Way 
\citep{Kannan:2014, Brook:2014, Dutton:2015, Stinson:2015, Tollet:2015, Wang:2015}.
The fact that these galaxies follow the the stellar mass - halo mass relation over a wide range in mass and redshift 
makes this sample a useful test for limiting star formation using stellar feedback.

Despite these successes, open questions still remain. In this study we focus on two of them: 
the paucity of disc dominated galaxies in simulations and the excess thickness of such discs.
\citet{Scannapieco:2010} suggested that some of the ``difficulty''
simulating bulgeless galaxies comes from inappropriately comparing observations and simulations.
Observations of ``bulgeless'' galaxies often rely upon surface density profiles.
Simulations consider the kinematics when determining the bulge fraction.
One consequence of using energetic stellar feedback to limit star formation is that the motions 
of gas are increased in simulations.
These increases lead to unrealistically thick discs \citep{Agertz:2013, Roskar:2014}.

Decomposing simulated galaxies is more straightforward than observed ones since we have the full six-dimensional (6D) phase information 
about the positions of particles and their velocities, as well as ages and abundances.   
All this information can also be used to create mock images of simulated galaxies based on stellar population models. 
Thus, we can make a comparison between photometric and kinematic decomposition techniques. 
Comparing these techniques can shed light on how galaxies formed and provide another test to see if galaxies simulated 
with stellar feedback share the properties of observed galaxies.

The current standard kinematic decomposition of simulated galaxies relies on analyzing the distribution 
of stellar circularities, $\epsilon$
\citep{Abadi:2003, Brooks:2008, Scannapieco:2010, Scannapieco:2011, Brook:2012, Martig:2012, Kannan:2015, Zavala:2015}. 
The circularity, $\epsilon$, is computed as $\epsilon(E) \equiv J_{\rm z}/J_{\rm c}(E)$, 
where $J_{\rm z}$ is the azimuthal angular momentum of a particle and 
$J_{\rm c}(E)$ is the angular momentum of a circular orbit having the same binding energy, $E$.
Such kinematic decompositions result in galaxies with higher bulge to disc ratios than photometric decompositions 
of observed galaxies typically find \citep{Scannapieco:2010}.  
Classic decomposition methods, however, make ambiguous classifications of stars with intermediate values of circularities.  

If we wish to trace the origins of the disc and bulge components, 
it would be helpful to eliminate, or at least minimize, these ambiguous classifications. 
Since $J_{\rm z}$ is only one dimension of 6D phase space, 
\citet[][hereafter DM12]{Domenech:2012} suggested adding binding energy, $E$, and the angular momentum component perpendicular 
to the disc, $J_{\rm p}$, to help classify stars less ambiguously. 
Also, instead of adopting fixed values in the various coordinates to discriminate between the two components 
\citep{Brooks:2008, Brook:2012}, they used a statistical cluster finding algorithm to let the data decide where it should best be split. 

In this paper, we use a sample of eighteen of \citet{Wang:2015}'s high resolution, ``zoom-in'' cosmological simulations 
from the \textit{Numerical Investigation of a Hundred Astrophysical Objects} ({\sc nihao}) project. 
We propose and test a new kinematic decomposition method based on the lines put forward by DM12.
The first aim of this study is to have a robust definition of discs in simulations, so that we can study their formation histories. 
The second is to compare observations and simulations in a more robust way, 
particularly the open issues of disc thickness and disc dominance.
Finally, having the sample of simulated galaxies spanning an order of magnitude in stellar mass allows us to look for 
trends with mass that can be subsequently used for predictions. 

The paper is structured as follows. Section \S \ref{sim_section} describes the simulation code and the galaxy sample. 
The kinematic decomposition method is presented in Section \S \ref{methods_section}. 
All the five properties that differentiate kinematic discs from kinematic spheroids are analysed in the results section, \S \ref{results}.
In the same section we also compare some of the properties of the simulated galaxies with observational data. 
The conclusions are presented in Section \S \ref{conclusions}.

\begin{figure*}
\includegraphics[width=1.0\textwidth]{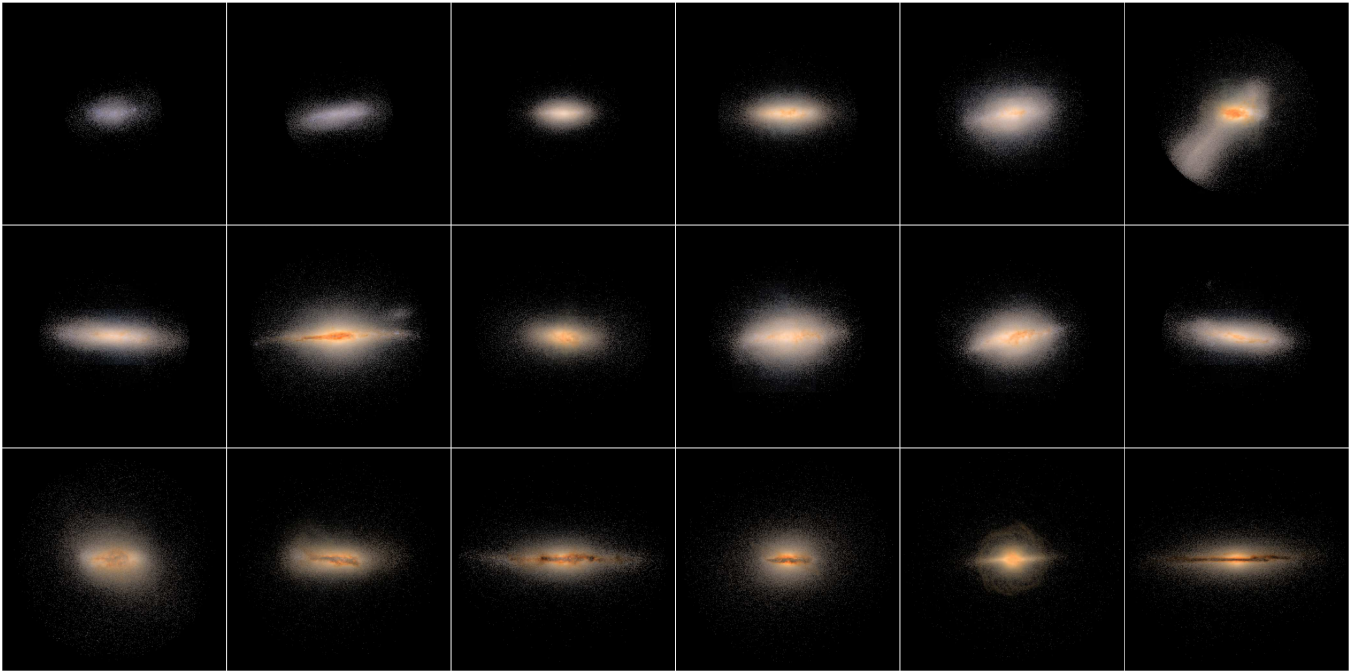}
\caption{Edge on composite images of the subsample of {\sc nihao}. The mass increases from left to right and top to bottom.
The galaxies have been aligned with the angular momentum of their gaseous discs. 
For this reason some of the stellar discs appear misaligned.}
\label{fig:sunrise}
\end{figure*}

\section{Simulations}
\label{sim_section}

The simulations we analyse in this work are a subsample of the largest high resolution, ``zoom-in'' 
cosmological simulation sample to date, {\sc nihao} \citep{Wang:2015}. 
These galaxies have been simulated with the new version of the N-body SPH code, {\tt GASOLINE} 
\citep{Wadsley:2004, Keller:2014, Keller:2015},
assuming the latest Planck cosmology (Planck Collaboration 2014). 

\subsection{Hydrodynamics}
The new version of the {\tt GASOLINE} code has several improvements with respect to the older versions. 
The hydro-solver has been modified following \citet{Ritchie:2001}
so that the problem of the artificial cold blobs was greatly alleviated. 
One of the important differences regards the use of a different SPH kernel \citep{Dehnen:2012} 
and an increased number of neighbors (now 50). 
Other relevant changes in the hydro-solver are the treatment of artificial viscosity, now done as described by \citet{Price:2008}, 
and the use of the time step limiter proposed by \citet{Saitoh:2009}. 

Metal diffusion has been implemented as described in \citet{Wadsley:2008}. 
The sources contributing to the heating function are photoionization and photoheating by a constant UV background \citep{Haardt:2012},
while the cooling function includes the effects of metal lines and Compton scattering \citep{Shen:2010}. 

\subsection{Star Formation and Feedback}
Star formation follows a Kennicutt-Schmidt relation, where the eligible gas has a temperature lower than 15000 K and 
a density higher than 10.3 cm$^{\rm -3}$. Supernova feedback is implemented following the blast-wave formalism \citep{Stinson:2006}. 
Another stellar feedback mechanism ejects energy prior to supernovae explosions \citep{Stinson:2013a}.  
It accounts for the photoionizing radiation of massive stars prior to their SN phase. 
The code traces the evolution of heavy elements produced by SNe Ia \citep{Thielemann:1986} 
and SNe II \citep{Woosley:1995}, assuming a Chabrier IMF \citep{Chabrier:2003}. 
The free parameters in the feedback scheme 
have been chosen so a Milky Way mass galaxy fit the halo mass -- stellar mass relation at $z=0$ \citep[see][]{Stinson:2013a}.

\subsection{Sample}
\label{sample}
The halo selection for {\sc nihao} ``zoom-ins'' used the cosmological dark matter only simulations presented in \citet{Dutton:2014}, 
without considering merger history. The {\sc nihao} galaxies cover a halo mass from 10$^{\rm 9.7}$ to 10$^{\rm 12.3}$ M$_{\odot}$. 
The simulations include around $10^6$ dark matter and $10^6$ gas particles inside $r_{\rm rvir}$.
The galaxies span two resolution regimes on either side of $5.55\times10^{11}$ M$_{\odot}$ 
that are described in detail in \cite{Wang:2015}.
At lower masses, the gravitational softening for star particles is 200 pc; at higher masses, it is 400 pc.

We selected a subsample of galaxies from {\sc nihao} based on their edge-on discy appearance in composite {\sc sunrise}
\citep{Jonsson:2006, Jonsson:2010} images.  The {\sc sunrise} images are shown in Fig.~\ref{fig:sunrise}. 
These galaxies all lie in the halo mass range [10$^{\rm 11}$--10$^{\rm 12}$M$_{\odot}$]. 
The resulting sample of eighteen objects includes galaxies spanning a whole decade in halo and stellar mass. 

\section{Decomposition technique}
\label{methods_section}

Kinematic decomposition of stellar systems has evolved over the last decade.
\citet{Abadi:2003} introduced a one dimensional decomposition 
based on the circularity of stellar orbits, a method we refer to as ``classic''.
DM12 added two more dimensions,
total binding energy and motions perpendicular to the disc plane. 
They used the \textit{k-means} algorithm to classify particles.
We present a refinement in which we use Gaussian Mixture Models for particle classification. 

\subsection{The classic method}

The classic method of \citet{Abadi:2003} does a good job determining the relative mass of disc and spheroid
\citep[see also][]{Scannapieco:2011, Martig:2012}. 
It assumes that the spheroid is non-rotating and thus has a symmetric distribution of circularities around $\epsilon$=0. 
The thin disc gives the sharp peak at circularities close to maximum  $\epsilon$=1.
Any extra features not accounted for by the thin disc and spheroid are considered to be driven by thick discs. 

This decomposition method assigns particles with positive circularities to one component or another,
such that the resulting probability distribution for the spheroid (no net rotation) is symmetric. 
This assignation, however, is not unique. 
Thus, it is difficult to study the properties of discs and spheroids individually, or to follow their evolution.

\subsection{k-means in multidimensional spaces}

DM12 increased the number of dynamical parameters considered for the decomposition. 
They used an unsupervised clustering method in the multi--dimensional space to assign particles to one of three possible classes: 
thin discs, thick discs and spheroids. 
The two new parameters are the binding energy, $E$, and the in--plane component of the angular momentum normalized to $J_{\rm c}(E)$, $J_{\rm p}$/$J_{\rm c}(E)$. 
The motivation for considering the binding energy as a prior is that
discs are less bound than the primary component of the spheroid, the bulge. 
$J_{\rm p}/J_{\rm c}$ quantifies the amount of motion out of the disc plane.

\begin{figure}
\includegraphics[width=0.5\textwidth]{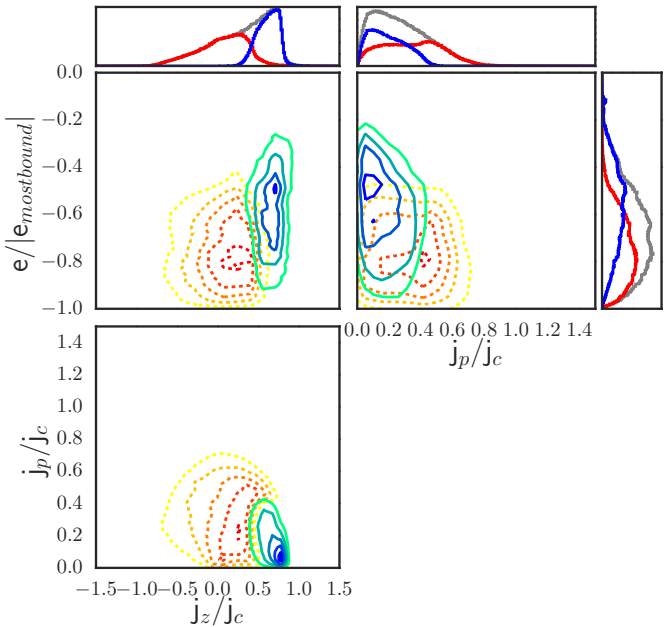}
\caption{The result of GMM applied to the galaxy g7.55e11. 
The three large panels give the 2D distributions as number density contours equally spaced, with solid green-blue and dashed yellow-red colors 
for the particles identified as belonging to the disc and spheroid, respectively.
The 1D distributions for $j_{\rm z}/j_{\rm c}$ and $j_{\rm p}/j_{\rm c}$ are shown in the left and right upper most panels, 
while for $e$/$|e|_{\rm mostbound}$ in the right upper panel horizontally oriented. In the 1D distributions, 
blue, red and grey stand for disc, spheroid and all stellar particles, respectively.}
\label{figure_3D_deco}
\end{figure}

The clustering method DM12 used to classify particles is \textit{k-means} \citep{Scholkopf:1998}.  
\textit{k-means} minimizes the intra--cluster distance,
where the intra--cluster distance is the sum of squared distances for all cluster pairs, 
also referred to as the ``inertia'' criterion. 
The only input it requires is the number of clusters. 
The advantage of \textit{k-means} is that it always converges, given enough iterations. 
However, the solution might be a local minimum.
The number of iterations can be significantly reduced if the initial centroids are chosen carefully 
(e.g. as far away from each other as possible or based on prior information on the data set) instead of being randomly assigned. 
The main disadvantage stems from two underlying assumptions that clusters are convex and isotropic: 
\textit{k-means} behaves poorly on irregular shaped manifolds or non-spherical clusters. 
Another disadvantage is that the ``inertia'' criterion implies that clusters will tend to have comparable number of members.
This means the method will only be suitable for galaxies having the mass roughly equally distributed between the various components. 

This method was applied to a small sample of simulated galaxies by \citet{Obreja:2013} to study the properties 
of the galactic bulge components. 
\citet{Dominguez:2015} used it to study the evolution of the stellar disc and spheroid material within the Large Scale Structure.

The simulated galaxy sample we analyse here is larger and covers a wider range in galaxy morphologies 
than the objects in DM12, \citet{Obreja:2013} or \citet{Dominguez:2015}. 
For these reasons, the assumptions \textit{k-means} uses (clusters convexity, isotropy and similar weights) 
could lead to important biases. 

\subsection{Gaussian Mixture Models}

We replace \textit{k-means} with \textit{Gaussian Mixture Models} (GMM) clustering method 
that does not suffer from the above mentioned drawbacks.

GMM is probabilistic in nature and assumes 
data points are drawn from a mixture of Gaussian distributions with unknown means and variances. 
Similar to the \textit{k-means} method, it uses an expectation--maximization algorithm to find the parameters of the Gaussians. 
Unlike \textit{k-means} it allows the data to have various covariance structures. 
The metric used is the Mahalanobis distance to the cluster centres (means of the Gaussians). 
The clustering algorithm on the simulated sample has been run with the Python package for Machine Learning, 
{\tt scikit-learn} \citep{Pedregosa:2011}. 
The Python code that does the decomposition is now available upon request. 
It will be publicly available by the end of this year.
The only input it needs is a simulation snapshot which is read using the {\tt pynbody} package \citep{pynbody}.

We use the same parameter space of ($j_{\rm z}/j_{\rm c}$,$j_{\rm p}/j_{\rm c}$,$e$), 
except that we switch to \emph{specific} angular momentum and binding energy, for which we use lower case letters. 
We also scale the binding energies of stars in each galaxy from -1 to 0, 
such that -1 represents the most bound star particle and 0 represents the least bound star particle. 
This scaling factors out the galaxy/halo mass dependence and removes the dimensionality of the energy. 

In this first study of galaxy structure we limit ourselves to run the clustering algorithm with only two components. 
The reason to do so is that in this case GMM is able to clearly distinguish the disc, as we will show. 
From now on, we will be referring to the non-disc component as spheroid.  

As an example, Fig.~\ref{figure_3D_deco} shows the full 3D clustering analysis for one galaxy projected onto the three axes, 
as well as the cumulative distributions for each dimension. 
The uppermost panel in the left column shows the classic 1D circularity distributions for 
particles classified as disc/spheroid in blue/red, as well as for all particles in grey. 
This galaxy shows a disc component that spans circularities between 0.3 and 1.0. 
The maximum of the circularity distribution is the disc component at $\sim$0.9. 
The spheroid component is almost symmetric and centered at $\sim$0.0. 
The spheroid's right wing breaks the symmetry as it increases while the left one monotonically decreases. 
The $\epsilon$ distributions for the disc and spheroid overlap 
over an extended range (from $\sim$0.3 to $\sim$0.8), 
but the other projections make it clear why particles were assigned to each component.  

The centre-left panel shows the binding energy as a function of $j_{\rm z}/j_{\rm c}$. 
In it, the disc and spheroid particles mostly occupy different regions, though there is some overlap for the most bound particles. 
The lower-left panel that shows $j_{\rm p}/j_{\rm c}$ as a function of $j_{\rm z}/j_{\rm c}$ clarifies that the overlap
comes from particles that have significant motions out of the disc plane. 
These particles with high perpendicular velocities are assigned to the spheroid.
For the rest of this work we will be referring to the two galaxy components as disc and spheroid. 
We make no attempt to divide the spheroid into bulge and halo.   

\section{Results}
\label{results}

Using the GMM decomposition, we kinematically separate discs from spheroids in the simulation sample.  
As a first step, we compare the GMM method with the classic one.
Next we study all the parameters that differentiate discs from spheroids: 
intrinsic shape, specific angular momentum, rotational velocities, 
velocity dispersions, ages, abundances and surface mass density profiles.  
The main reason to do so is to test if the various disc definitions are consistent.
All the quantities analysed are mass weighted, and as a final step, we make some comparisons with observations. 
The large sample of galaxies can uncover trends in disc properties that can be tested against observations.

Throughout this study discs are represented as stars, $\star$ and spheroids as circles, $\circ$. 
In most cases, the spheroids are colored red and the discs blue. 
In some cases, the simulation data points are colour coded by various quantities to point out the correlation between parameters.
The parameters derived for all the stars in the galaxy are given as empty (or filled) black squares.
Observational data are shown as grey symbols. 

\subsection{Kinematic decomposition of NIHAO}
\label{KD-nihao}

\begin{figure*}
\includegraphics[width=0.98\textwidth]{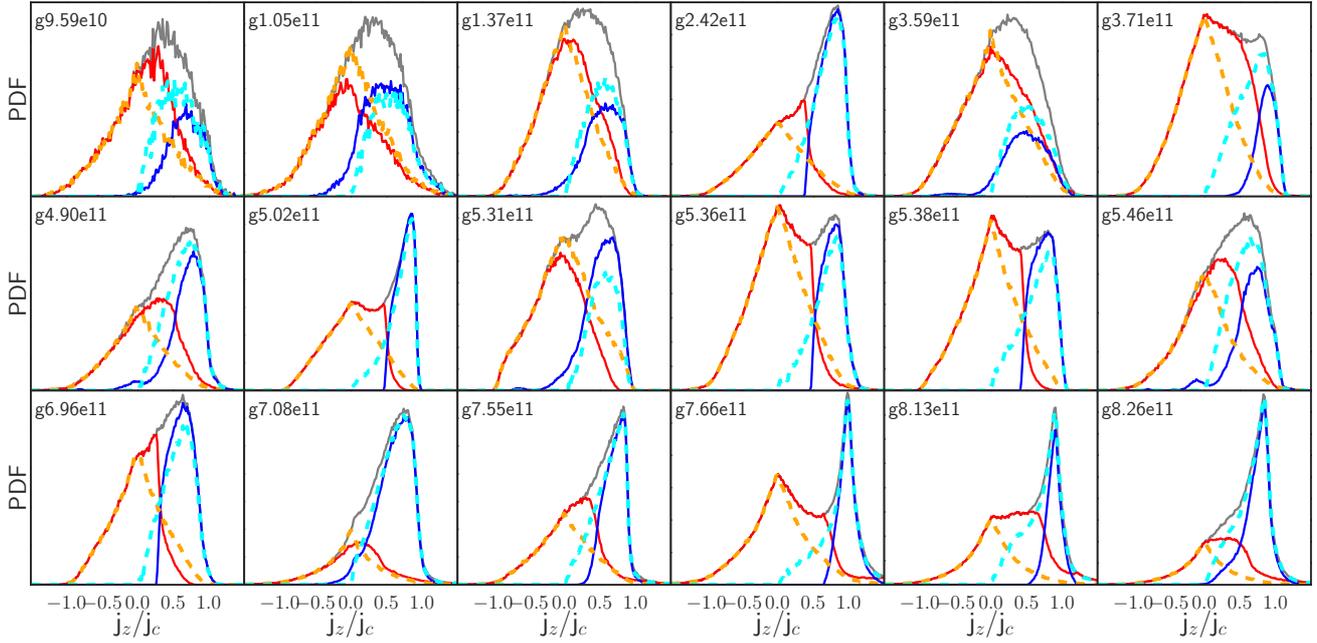}
\caption{The results of the two kinematic decompositions on the eccentricity parameter j$_{\rm z}$/j$_{\rm c}$ for the galaxies in the sample.
Solid grey, red and blue curves stand for the full galaxies, the GMM spheroids and the GMM discs, respectively. 
Dashed orange and cyan curves stand for the spheroids and discs decomposed with the classic method.}
\label{figure11}
\end{figure*}

For each of the eighteen selected galaxies, we used the GMM algorithm to assign all possible stars inside $r_{\rm vir}$ 
to one of two components, disc or spheroid.  
Table \ref{table1} gives some global parameters of the sample based on their decomposition: 
halo mass (M$_{\rm h}$), stellar mass (M$_{\rm *}$), virial radius (r$_{\rm vir}$), circular velocity at 0.1r$_{\rm vir}$ (v$_{\rm c}$),
and disc-to-total ratios (D/T) from the GMM and classic kinematic decompositions, 
using an empirical expression derived from IFU data by \citet[][hereafter RF12]{Romanowsky:2012},
and from the 2D {\tt Galfit} \citep{Peng:2002, Peng:2010} decomposition of the face-on \textit{i}-band {\sc sunrise} images.

\begin{table*}
\centering
\begin{tabular}{ccccccccc}
\hline
Sim & M$_{\rm h}$ & M$_{\rm *}$ &r$_{\rm vir}$ & v$_{\rm c}$ & $D/T$ & $D/T$ & $D/T$ & $D/T$\\
  & 10$^{\rm 11}$M$_{\odot}$ & 10$^{\rm 9}$M$_{\odot}$ & kpc & km s$^{\rm -1}$ & GMM & Classic & RF12 & Galfit\\
\hline
g9.59e10 & 0.88 & 0.24 & 95 & 68 & 0.18 & 0.38 & 0.30 & 0.99\\
g1.05e11 & 1.18 & 0.51 & 105 & 80 &  0.40 & 0.33 & 0.32 & 1.00\\
g1.37e11 & 1.48 & 1.99 & 112 & 101 & 0.28 & 0.34 & 0.29 & 1.00\\
g2.42e11 & 2.68 & 5.36 & 136 & 127 & 0.53 & 0.59 & 0.51 & 1.00\\
g3.59e11 & 3.49 & 4.16 & 150 & 120 & 0.17 & 0.29 & 0.26 & 1.00\\
g3.71e11 & 4.08 & 11.14 & 159 & 141 & 0.05 & 0.31 & 0.38 & 0.95\\
g4.90e11 & 3.16 & 2.83 & 146 & 125 & 0.47 & 0.62 & 0.55 & 0.99\\
g5.02e11 & 5.75 & 14.32 & 176 & 157 & 0.41 & 0.46 & 0.47 & 0.85\\
g5.31e11 & 5.28 & 16.25 & 172 & 156 & 0.39 & 0.28 & 0.27 & 0.97\\
g5.36e11 & 6.09 & 11.52 & 188 & 153 & 0.58 & 0.24 & 0.33 & 0.86\\
g5.38e11 & 6.31 & 18.25 & 184 & 166 & 0.48 & 0.54 & 0.57 & 0.98\\
g5.46e11 & 3.15 & 3.61 & 147 & 124 & 0.32 & 0.51 & 0.46 & 1.00\\
g6.96e11 & 7.86 & 32.32 & 197 & 185 & 0.49 & 0.47 & 0.39 & 1.00\\
g7.08e11 & 8.01 & 35.25 & 197 & 197 & 0.56 & 0.58 & 0.54 & 0.97\\
g7.55e11 & 8.22 & 30.26 & 204 & 196 & 0.45 & 0.57 & 0.55 & 0.60\\
g7.66e11 & 9.30 & 55.21 & 207 & 208 & 0.30 & 0.45 & 0.48 & 0.83\\
g8.13e11 & 9.91 & 62.61 & 211 & 218 & 0.29 & 0.63 & 0.58 & 0.97\\
g8.26e11 & 10.21 & 45.17 & 213 & 222 & 0.56 & 0.75 & 0.67 & 0.67\\
\hline
\end{tabular}
\caption{Simulation, halo mass (M$_{\rm h}$), stellar mass (M$_{\rm *}$), virial radius (r$_{\rm vir}$), circular velocity at 10\% of the virial radius (v$_{\rm c}$), 
and disc-to-total ratios (D/T) from the GMM and the classic kinematic decompositions, 
computed using the expression of RF12, and from the 2D {\tt Galfit} decomposition of the face-on \textit{i}-band {\sc sunrise} images for the sample of simulated galaxies.}
\label{table1}
\end{table*}

For a graphical summary of the decomposition, 
Fig.~\ref{figure11} shows the eccentricity distribution for every galaxy in our sample.
It compares the GMM decomposition (solid red is spheroid, disc is solid blue) with the classic method (dashed orange is spheroid, disc is dashed cyan).
The simulated galaxies are ordered based on their mass which increases from left to right and top to bottom. 
There are a couple of notable trends with mass. 
The low mass galaxies only have one peak in their PDFs. 
At higher masses two peaks become more evident. As the mass increases, the higher peak also increases to greater circularity. 
Thus, higher mass galaxies have a higher fraction of stars moving on exactly circular orbits.
This translates in less massive systems having more homogeneous stellar populations characteristic of spheroids, 
while more massive systems are more disc dominated. 

In the GMM decomposition, some discs show a small fraction of counter-rotating stars. 
Counter-rotating stars are allowed in the GMM method because their parameters are closer
to the disc component than the spheroid.  They are typically stars near the centre of the galaxy.
Their minimal presence supports the choice of not using priors for the decomposition. 

The decomposition explained for the galaxy shown in Fig.~\ref{figure_3D_deco} is typical, but 
no two galaxies look exactly the same in this parameter space. In their 1D distributions, the galaxies show common features: 
\begin{enumerate}
\item the maximum of the circularity distribution is always positive indicating that every galaxy has a net rotation, 
\item the spheroid includes stars with the highest j$_{\rm p}$/j$_{\rm c}$ values; the disc j$_{\rm p}$/j$_{\rm c}$ distribution is skewed towards 0.0
\item the disc is less bound than the spheroid and covers a larger range in binding energies.  
\end{enumerate}
Regarding the 2D distributions, a clear trend with total stellar mass emerges in the sense that in the j$_{\rm z}$/j$_{\rm c}$ -- j$_{\rm p}$/j$_{\rm c}$ plane
there is less and less overlap between discs and spheroids with increasing mass, 
basically reaching no-overlap for the most massive galaxies. 

\begin{figure*}
\includegraphics[width=0.98\textwidth]{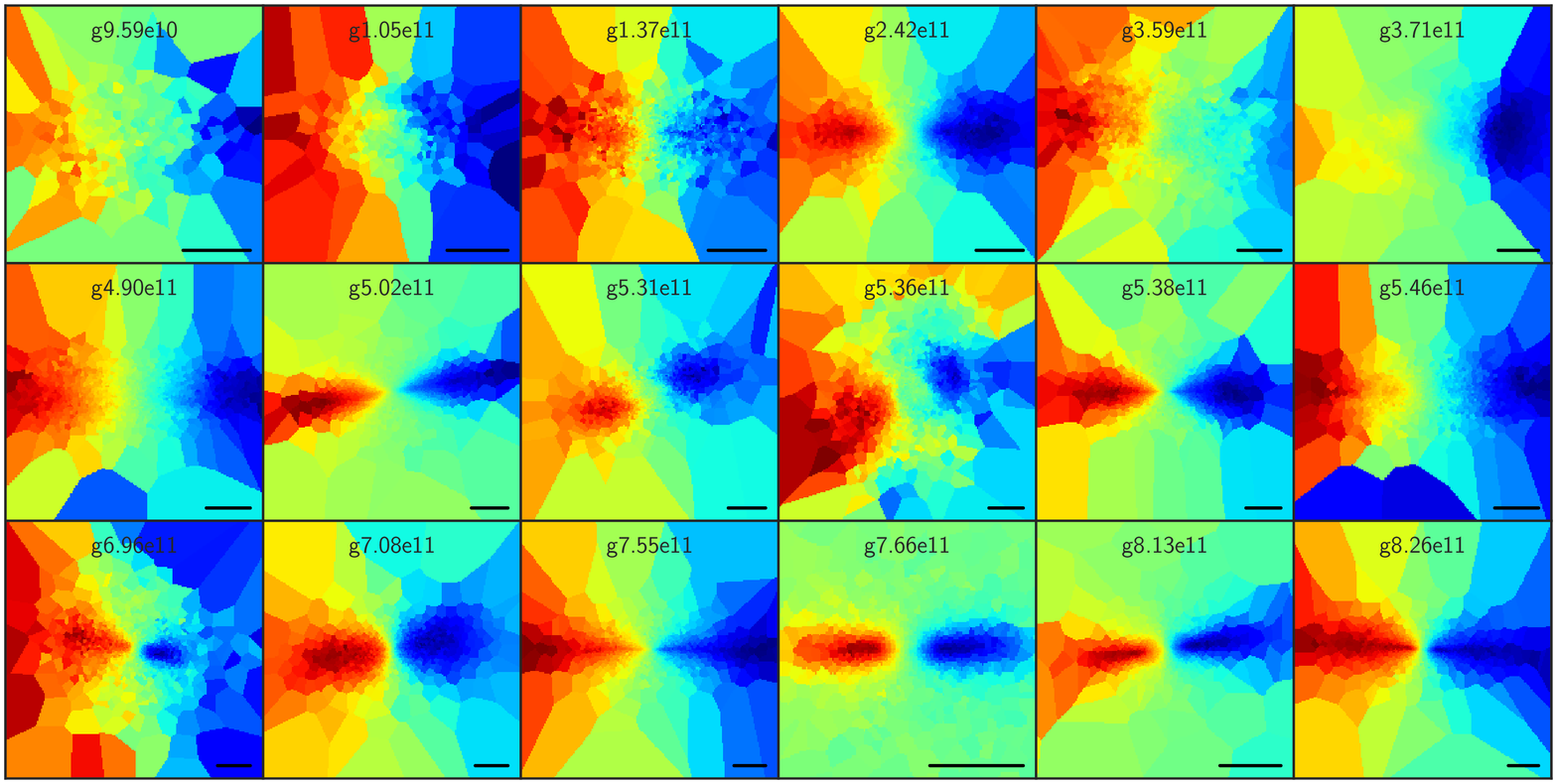}
\caption{The mock line of sight velocity fields in edge-on perspective for the eighteen galaxies in the sample. The color scale has been adapted separately for each galaxy
to extend from -v$_{\rm max}$ to v$_{\rm max}$, where the maximum has been taken over all the spaxels in the map. 
The horizontal black lines represent the physical scale of 5~kpc.}
\label{sample_vlos}
\end{figure*}

For the galaxies with two distinct maxima in the circularity PDF, demanding two components makes sense.
However, the others exhibit a broad single peak, so the choice is less clear.  
The width of the j$_{\rm z}$/j$_{\rm c}$ distribution suggests the component is a spheroid, but
the median of the peak is always positive, indicating net rotation.  It is thus possible
that low mass galaxies are fast rotating spheroids.  Fig.~\ref{sample_vlos} shows
mock IFU data for the simulations.  These kinematic signatures are much more similar to discs.
The method of constructing the IFU maps is explained in detail in the Appendix. 
Thus, our current analysis is inconclusive on whether the decomposition should
be limited to less than two components.

Whether more than two components should be allowed is another concern.
The current method cleanly defines thin discs, which are the focus of the present study.
Therefore, attempts to search for more than two components are deferred to future work.

\subsection{Disc--to--Total Ratios}

\begin{figure}
  \includegraphics[width=0.5\textwidth]{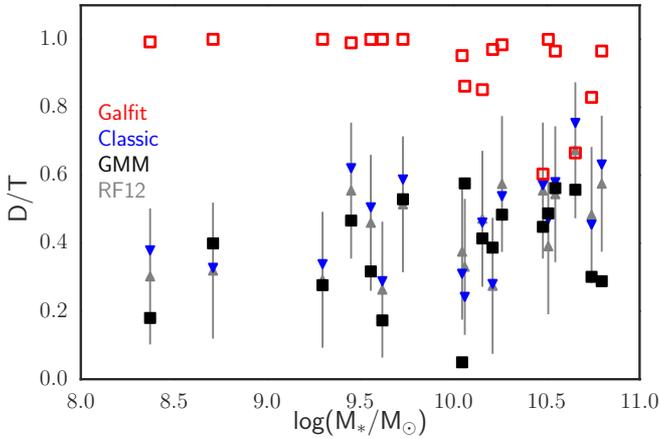}
  \caption{Disc-to-total ratios as a function of total stellar mass.
  The results of the classic kinematic decomposition of \citet{Abadi:2003} are shown as blue triangles, 
  those of the GMM method explained in Section~\ref{methods_section} as filled black squares, 
  while grey triangles give the values computed following RF12, as explained in the text.
  The empty red squares represent the photometric ratios derived from the {\sc sunrise} \textit{i}-band images using {\tt Galfit}.}
 \label{fig_mass_dt}
\end{figure}

The fundamental result of the decomposition is how much mass is in each component. 
Fig.~\ref{fig_mass_dt} shows the fraction of the total mass classified into the disc component (D/T) as a function of stellar mass. 
The GMM and classic D/T ratios are shown as the filled black squares, $\blacksquare$, and filled blue triangles, $\blacktriangledown$. 
They have roughly the same values for each galaxy, with the classic decomposition yielding 5-20\% higher values in nearly every case. 
The small difference between the kinematic decompositions stems from the classic method assuming a symmetric spheroid (no net rotation), 
while the GMM method has a surplus of positive circularity stars in every galaxy.
Therefore, D/T for the GMM method is systematically smaller than D/T for the classic one.

While it is hard to observationally decompose galaxies based on kinematics, IFU are enabling initial attempts. 
Zhu et al. in prep. are applying Schwarzschild models to spiral galaxies IFU data. 
In this way various components of observed galaxies are separated by classifying their reconstructed stellar orbits.
As a first step, RF12 proposed a simpler decomposition to quantify D/T based on a galaxy's observed kinematics. 
Assuming discs and bulges have exponential and de Vaucouleurs profiles, 
they define the kinematic disc fraction as:

\begin{equation}
 D/T = 1.35v/\sigma\sqrt{2.+(1.35v/\sigma)^2},
\end{equation}
where the factor 1.35 comes from projection effects and $v/\sigma$ is summed over a galaxy from IFU spaxels such that:

\begin{equation}
v/\sigma \equiv \sqrt{\frac{\Sigma_i f_iv_{los,i}^2}{\Sigma_i f_i\sigma_{los,i}^2}}.
\label{eqv2sigma}
\end{equation}
In Equation~\ref{eqv2sigma}, $i$ runs over the spaxels inside an elliptical aperture of area $\pi R_{\rm e}^{\rm 2}$,  
where R$_{\rm e}$ is the half light radius. f$_{\rm i}$ is the flux, v$_{\rm los,i}$ the line of sight velocity 
and $\sigma_{\rm los,i}$ the line of sight velocity dispersion of spaxel $i$. 

Using the velocity moments maps constructed as explained in the Appendix, 
we compute v/$\sigma$ for the simulated galaxies. 
Since we are concerned with the intrinsic properties of the simulated galaxies, 
the spaxels contributions are weighted by mass instead of light. 
The values of v/$\sigma$ computed within half mass radii are low, so D/T is low ($\sim$0.2). 
If v/$\sigma$ is computed to include the entire galaxy, R=0.1r$_{\rm vir}$, instead, the values become comparable
to those found using the kinematic decomposition.
These values are given as grey triangles, $\blacktriangle$, in Fig.~\ref{fig_mass_dt}. 
While our calculation does not follow the observed measure exactly, it shows that as kinematic observations improve, 
the results should converge to what is found with complete kinematic information as we have from the simulations.
However, it is also a useful result since it proves an important point. 
If we were able to have high S/N kinematic 2D data at large radii, we could in principle recover true dynamical fractions. 
The difference in D/T found from the central region and the entire galaxy is unsurprising as the centre is bulge dominated.

In the same figure the results of photometric decompositions are also provided as the empty red squares, $\square$.
For the photometric decompositions, mock face-on $i$-band images were created using the Monte Carlo radiative transfer code, {\sc sunrise}.  
{\sc sunrise} calculates radiative transfer using a Monte Carlo scheme to take into account the effects of dust. 
The publicly available program {\tt Galfit}
performs a least-squared fitting routine to determine 2D galactic structural parameters. 
To determine the photometric disc-to-total ratio, we first fit the galaxy with a pure exponential disc. 
We use the results from the pure exponential fit as starting parameters for a two component fit 
consisting of an exponential disc with a Sersic index of 1 and a de Vauccouleurs bulge with a Sersic index of 4. 
In Fig.~\ref{fig_mass_dt}, the D/T that we report is L$_{\rm disc}$ / (L$_{\rm disc}$ + L$_{\rm bulge}$).

A comparison of the kinematic and  photometric decompositions shows a significant discrepancy between the galaxy type inferred
\citep[see also][]{Scannapieco:2010}. 
The disc-to-total ratios obtained from 2D fits of the \textit{i}-band surface brightness 
are very close to 1 (89\% of galaxies have D/T$>$0.8). 
However, defining D/T as the fraction of stellar mass in the kinematic disc, the values obtained are around 0.4 with a big scatter. 
The highest kinematic D/T is 0.8 for only one galaxy, while all the rest have ratios smaller than 0.6.
The stellar surface mass density profiles of most of the galaxies in the sample tell the same story as the photometric D/T,
namely that they are in most cases pure exponentials. 
This result lessens concern about forming bulgeless discs in simulations. 
Forming pure exponentials in simulations is straightforward. 
However, those ``discs'' are not necessarily kinematically cold.
It also highlights the importance  of comparing simulations with observations using the same metric. 

Fig.~\ref{fig_mass_dt} highlights the large difference in D/T ratios found using kinematic versus photometric decompositions. 
Most of the low mass galaxies have exponential profiles, so the photometric decomposition
gives D/T$\cong$1 for all the low mass galaxies while the kinematic D/T ratios for both 
the GMM and the classic method result in ratios in the range [0.2, 0.6].  
As the photometric D/T ratios fall at higher masses, the kinematic D/T ratios rise.

\subsection{Kinematics}
\label{sec:kinematics}
To examine the properties of the individual kinematic components,
we begin our analysis with a comparison between two fundamental
properties of galaxies, their stellar mass and specific angular momentum.
\citet{Fall:1983} placed observed galaxies onto this plane and found
that $j_{\rm *}$-$M_{\rm *}$ separated galaxies into distinct morphologies.
In this way, the galaxy's spin parameter, $\lambda=J E^{\rm 1/2}/G M^{\rm 5/2}$,
should determine a galaxy's morphology.
Additionally, inside each galaxy morphology, the galaxies followed a 
relation of $j_{\rm *} \propto M_{\rm *}^{\rm 2/3}$.  For disc galaxies, that relation
is the same as what \citet{Tully:1977} found.  A simple unit based argument
for the relation is that
\begin{equation}
j \sim V R 
\end{equation}
\begin{equation}
V \sim (M/R)^{1/2} 
\end{equation}
\begin{equation}
R \sim M^{1/3} 
\end{equation}
\begin{equation}
j \sim M^{2/3}
\end{equation}
where j is the specific angular momentum, V is the velocity, R is the radius and M is the mass of the entire galaxy halo.  
RF12 followed up \citet{Fall:1983} with data that extended to larger radii. 
They found that there was not a significant amount of angular momentum in the outer regions of elliptical galaxies 
and that the $j_{\rm *} \propto M_{\rm *}^{\rm 2/3}$ still held.

\begin{figure}
\includegraphics[width=0.5\textwidth]{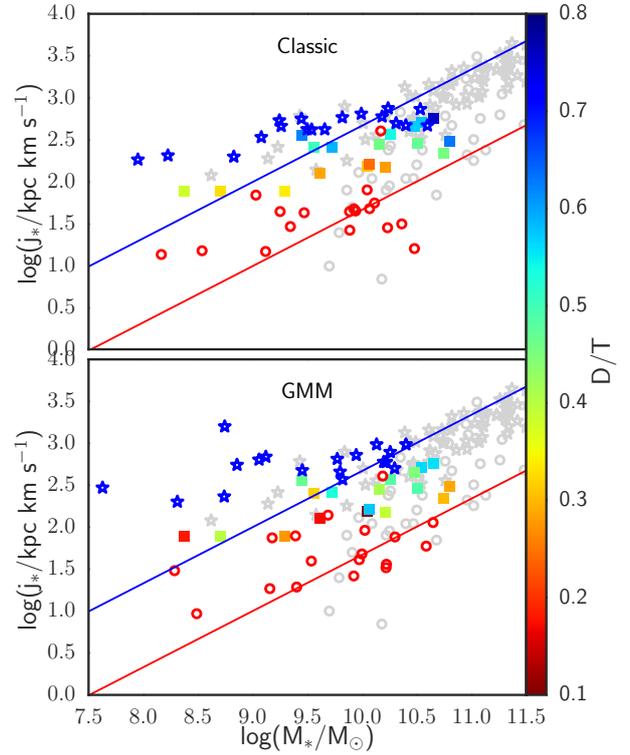}
\caption{Specific angular momentum as a function of stellar mass for the kinematically defined spheroids (red circles) and discs (blue stars). 
The red and blue lines give the theoretical predictions for pure bulges and pure discs from RF12,
while the grey stars and circles show the observational data sets of late types and ellipticals galaxies that they analysed. 
The simulated spheroids follow the pure bulges line, albeit with a large scatter, while the simulated discs also follow a power law, but
with a shallower slope than the theoretical prediction. 
The squares represent the full galaxies, color coded by their respective kinematically defined disc-to-total ratios. 
The upper panel gives the results obtained using the classic decomposition, while the bottom one those of the GMM.}
\label{fig_j_mass}
\end{figure}

In simulations, one can compute specific angular momentum as:

\begin{equation}
 j = \frac{|\Sigma_k m_k \overrightarrow{r_k} \times \overrightarrow{v_k}|}{\Sigma_k m_k},
\end{equation}
where m$_{\rm k}$, r$_{\rm k}$ and v$_{\rm k}$ are the mass, position and velocity of particle $k$, and $k$ ranges over all stellar particle associated
with the whole galaxy, the kinematic disc or the kinematic spheroid. All galaxies have been initially aligned such that the vertical 
direction is the perpendicular to the gaseous disc plane. 

Fig.~\ref{fig_j_mass} shows how the specific angular momentum varies as a function of mass 
for the simulated galaxies.  The disc (\emph{blue stars}) and spheroid (\emph{red circles}) 
components are shown separate from the total galaxy's.  
The total galaxies are given as squares colored according to their respective kinematic decomposition disc fractions.
Observed late type galaxies and ellipticals are shown as grey stars and circles, respectively. 
The red and blue lines use the same normalization as RF12, and both have a slope of 2/3. 

The simulated spheroids follow the pure bulge line, albeit with some scatter. 
In contrast, the discs follow a different power law from the theoretical prediction, one that is flatter.  
At low masses, the simulated disc components have more angular momentum than the RF12 prediction.
The data points giving the full galaxies also follow a power law, more similar to the disc, with a shallower slope than the 
observationally derived function. While the simulations differ somewhat from the theoretical predictions, 
it is clear that the discs have more rotational support than the spheroids. 
In any case, simulations overlap to a large extent with the observational data analysed by RF12. 

A couple spheroid components have specific angular momenta values close to theoretical discs.  
In this plane, the two kinematic decompositions give very similar results. 
The most notable difference, which is still minor, can be seen in the spheroid component. 
While GMM gives spheroids that follow closely the pure bulge function of RF12, 
the classic method results in shallower slope with stellar mass. 

\subsection{Disc thickness}
To form disc galaxies with the right stellar mass and flat rotation curves, 
simulations typically use stellar feedback to limit star formation
\citep[e.g.][]{DallaVecchia:2012, Stinson:2013a, Aumer:2013, Hopkins:2014}.
Stellar feedback drives turbulence in star forming gas.  
Simulations typically have larger scale heights 
and vertical velocity dispersions than observations \citep{Stinson:2013b, Roskar:2014}.  
In addition to feedback driven turbulence, resolution can be a concern.  
The scaleheight of the thin disc of the Milky Way that includes the youngest stars is $\sim$100 pc \citep{Larsen:2003}.
A Milky Way mass galaxy in {\sc nihao} has a spatial resolution of 400 pc, so it
is numerically impossible for the disc to be as thin as the Milky Way's thin disc.

To date, disc thickness measurements have only been presented for single simulated galaxies.  
\citet{Roskar:2014} simulated one galaxy using a variety of parameters.  
Comparing single galaxy disc thicknesses is a challenge because observed galaxy disc thicknesses
vary as a function of stellar mass \citep{Dalcanton:2002, Yoachim:2006}. 

\begin{figure}
 \includegraphics[width=0.5\textwidth]{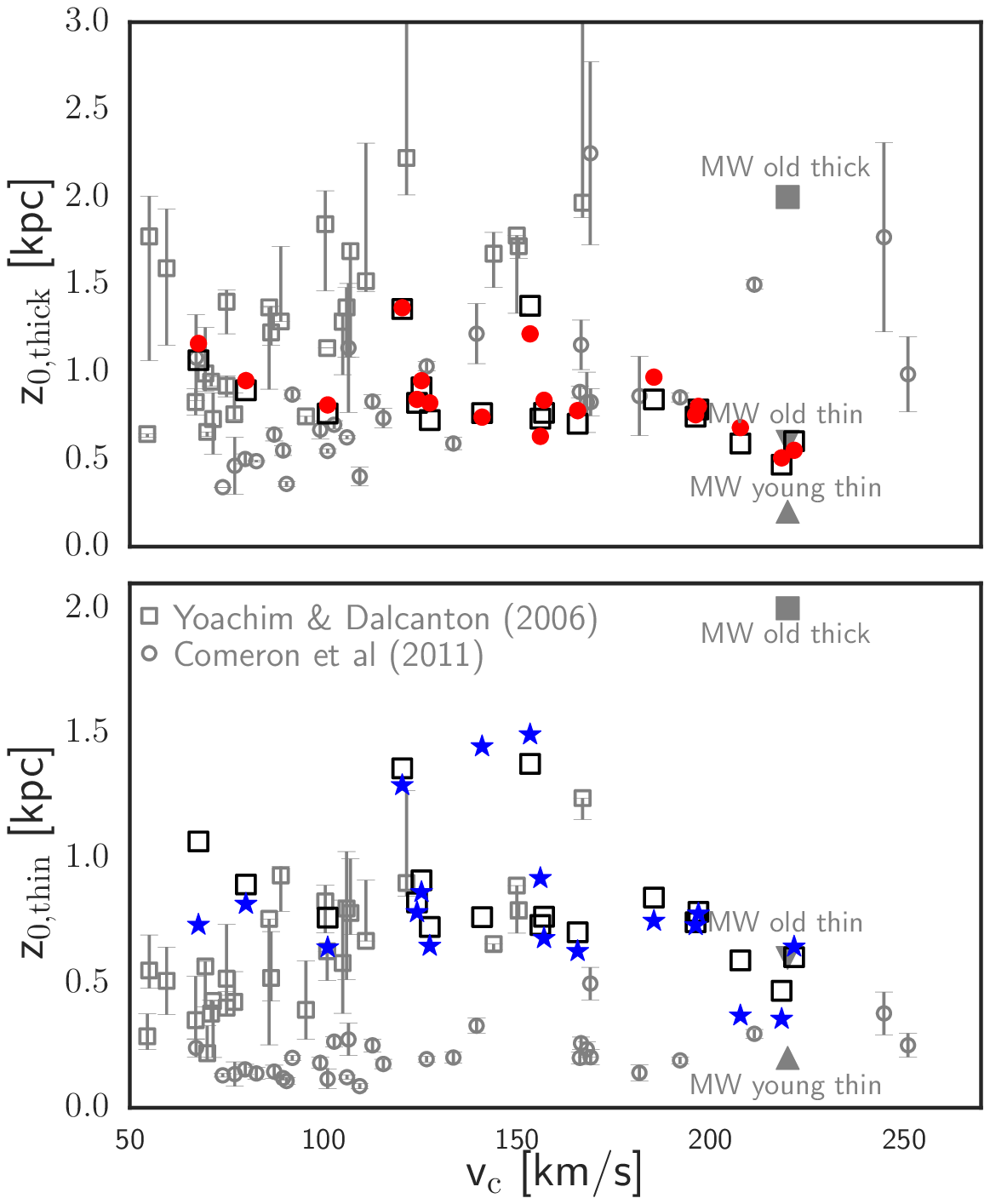}
 \caption{Disc thickness, z$_{\rm 0}$, as a function of circular velocity, v$_{\rm c}$, 
 for the spheroid (red circles in top panel) and disc (blue stars in bottom panel) components.  
 The simulated disc thicknesses are compared with observed thicknesses of edge-on galaxies shown in grey.  
 \citet{Yoachim:2006} are shown as circles with error bars, \citet{Comeron:2011} are shown as square with error bars.  
 The simulations fall in the same thickness range as observations, but with the opposite trend.
 The observations increase in thickness with ${\rm v_c}$ while the simulations decrease. 
 The filled grey points tagged as ``MW young thin'', ``MW young thick'' and ``MW old thick''
 are the scaleheights for the Galaxy from \citet{Larsen:2003}.}
 \label{fig_scale_height}
\end{figure}

Fig.~\ref{fig_scale_height} shows a comparison between our {\sc nihao}
sample and several sets of observations (\emph{grey}) of disc thickness.
The disc thickness observations come from \citet{Yoachim:2006} and \citet{Comeron:2011},
two surveys that take photometry of a number of edge-on disc galaxies.
Determining disc thicknesses is challenging because even in nearby galaxies, discs are
only a few arcsec thick, close to the resolvable limit from the ground or with Spitzer.
Galaxies with $v_{\rm c} > 120$ km s$^{\rm -1}$ have prominent dust lanes that block most of the light from the galactic midplanes.  

Besides using different telescopes, these two studies compute disc scale heights in different ways.  
\citet{Yoachim:2006} uses a three parameter fitting function that simultaneously determines
the galaxy's scale length and separate thin and thick disc scale heights.
They use an exponential for the radial profile and $sech^{\rm 2}$ for the vertical profile.
$sech^{\rm 2}$ is commonly used to fit vertical profiles since it is flattens close to 0, but is exponential away from 0.  
Most galaxies show $sech^{\rm 2}$ vertical profiles, though some are also straight exponentials.
Some small decisions in the \citet{Yoachim:2006} fitting procedure pushed disc thicknesses to slightly higher values.

\citet{Comeron:2011} break galaxies into four radial segments at fixed fractions of $r_{\rm 25}$ and provide separate thickenesses at each radius.
\citet{Comeron:2011}'s fitting procedures tend to push thicknesses to lower values.
Fig.~\ref{fig_scale_height} shows a median value of the four \citet{Comeron:2011} scale heights and the range as the error bars.
Since the observed galaxies are edge-on, it is straightforward to determine their rotation velocities from gas emission lines.  

The studies obtain disc thicknesses that vary up to a factor of 4.  
However, they find thin-to-thick mass ratios that are quite similar.
We include both sets of observations to give a sense of how small choices in fitting observations can cause big changes in outcomes.

In \citet{Yoachim:2006}, the observed disc thicknesses increase with circular velocity.
In \citet{Comeron:2011}, the disc thicknesses remain nearly constant as a function of circular velocity.

The simulation scale heights are best fits of a $sech^{\rm 2}$ function to the vertical mass profile between 0.5 kpc $< z <$ 3 kpc.
For simplicity, the simulations are only fit along the z-dimension.
We assume that the gas emission lines closely trace the gravitational potential, so the simulated
galaxies are placed at their corresponding asymptotic circular velocity,  $v_{\rm c} = \sqrt{GM/0.1r_{\rm vir}}$.
The thicknesses of the individual components are plotted on separate frames: 
the spherical components (\emph{red circles}) are plotted against the observed thick discs;
the disc components (\emph{blue stars}) are plotted against the observed thin discs;
the total galaxy fit (\emph{empty black squares}) as shown in both panels.

The simulations fall within the observed range of disc thicknesses.
The disc thicknesses tend to decrease with increasing circular velocity.
There is also not much difference between the fits of the disc and spheroid components.
Typically, the spheroids are thicker than the discs, but not in every case.
In both observational samples, thick discs are up to four times thicker than the thin discs.
Part of the discrepancy may be due to differences in the measurement methods between simulations and observations.
It is surprising that the spheroids have such small scale heights and seems to present a better picture of simulated galaxies than what has been shown in the past.  
The simulations are thicker than the Milky Way, but other observed galaxies also appear to be thicker.

\subsection{Velocity dispersion}
\label{dms_comparison}

Velocity dispersion is a complementary measurement of how kinematically hot galactic components are.
IFUs make it possible to observe velocity dispersions across entire galaxies. 
The \textit{DiskMass Survey}  \citep[DMS, ][]{Bershady:2010} is an IFU survey of 146 galaxies.  
\citet{Martinsson:2013} present velocity dispersions for a face-on subsample of the DMS.
Their 30 galaxies are disc-dominated systems with no significant asymmetries, bars, bulges, or interacting companions. 
The sample thus represents the kinematically coldest face-on discs in the DMS.

Outside of selection effects, comparing vertical velocity dispersions from simulations with the DMS is relatively straightforward since 
the DMS sample uses the width of stellar absorption lines. 
These should correspond almost exactly with the scatter of the line of sight velocities in every spaxel of mock IFU maps. 
We create tessellated velocity maps of the dispersion in face-on images.
Like \citet{Martinsson:2013}, we create a vertical velocity dispersion profile for each galaxy and fit the profile with an exponential function. 
The central value of that exponential is the reported vertical velocity dispersion (see Table~\ref{table2}).
The details of the calculations are presented in the Appendix.

Comparing the rotation velocities is less straightforward than the velocity dispersions because the typical inclinations in the observational sample are less than $\sim30^\circ$. 
\citet{Martinsson:2013} provide several estimated of the rotational velocity based on deprojections, which might depend on the exact inclination measured for the galaxy. 
As a more robust proxy for rotation, they also derive the rotation velocity by inverting the Tully-Fisher relation in the K-band.
The NIHAO galaxies also fit on a tight Tully-Fisher relation when the rotation velocity is measured on the rotation curve determined by the galactic potential, $v_{\rm c} = \sqrt{GM/R}$. 
Therefore, we use the circular velocity at 10\% of the virial radius, called $v_{\rm c}$ in Table~\ref{table1}, for our comparison with their data. 

\begin{figure}
 \includegraphics[width=0.5\textwidth]{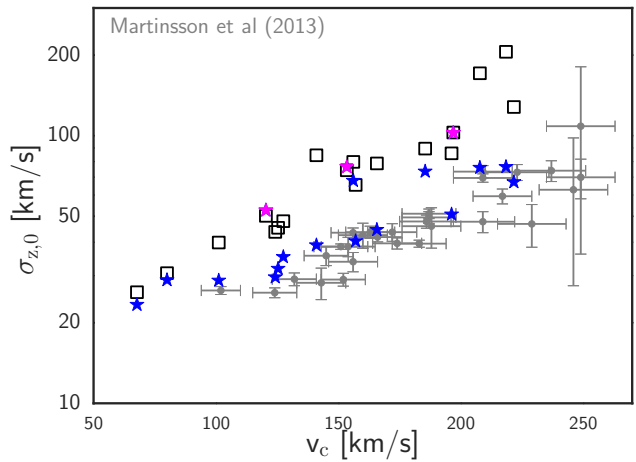}
 \caption{Vertical central line of sight velocity dispersion, $\sigma_{\rm z,0}$, as a function of asymptotic circular velocity, v$_{\rm c}$,
 for the full galaxies (empty black squares) and the GMM discs (blue stars). The magenta stars give the kinematic discs of g3.59e11, g5.36e11 and g7.08e11,
 whose peculiar behaviors are discussed in the text. 
 The grey squares show the observations by \citet{Martinsson:2013}.}
 \label{fig_velocity_sigma}
\end{figure}

Fig.~\ref{fig_velocity_sigma} shows vertical velocity dispersion as a function of rotation velocity. 
As might be expected for a selection of simulations being compared with a kinematically cold selection of observed galaxies, 
the vertical velocity dispersions of the total galaxies (empty black squares) are $\sim$2 times higher than the observations (grey squares). 
Such results are comparable with simulations run in the past.
Most of the kinematically selected discs (blue stars), however, fall on top of the observations. 
The kinematic decomposition is successful at extracting a kinematically cold component in most cases and that cold component is comparable to observed cold components in real galaxies.

The simulated discs sample shows a few outliers from the observational relation that we examine to find out whether anything is different about these galaxies. 
The three cases we examine are ones in which the vertical dispersion of the discs is equal to or exceeds the dispersion of the entire galaxy.
The simulation g3.59e11 results in the full galaxy and the kinematic disc (magenta star) having the same $\sigma_{\rm z,0}\sim$ 50 km s$^{\rm -1}$. 
The cause for this overlap is that the full galaxy is not well described by an exponential decreasing vertical velocity dispersion; 
$\sigma_{\rm z,0}$ in this case does not vary with the radius. However, the exponential decrease is a good approximation for the kinematic disc of the galaxy. 
The same situation happens with g5.36e11 which has $\sigma_{\rm z,0}\sim$ 75 km s$^{\rm -1}$ (its GMM disc is also shown in magenta).
These two galaxies show disturbed line of sight velocity maps (see Fig.~\ref{sample_vlos}).
The simulation g7.08e11, on the other hand, is clearly a  disc dominated system (see Figs.~\ref{fig:sunrise} and \ref{figure11}). 
In this case, this is the reason why the kinematic disc (magenta star) and the full galaxy have the same $\sigma_{\rm z,0}\sim$ 100 km s$^{\rm -1}$.

\subsection{Shapes}

\begin{figure}
 \includegraphics[width=0.5\textwidth]{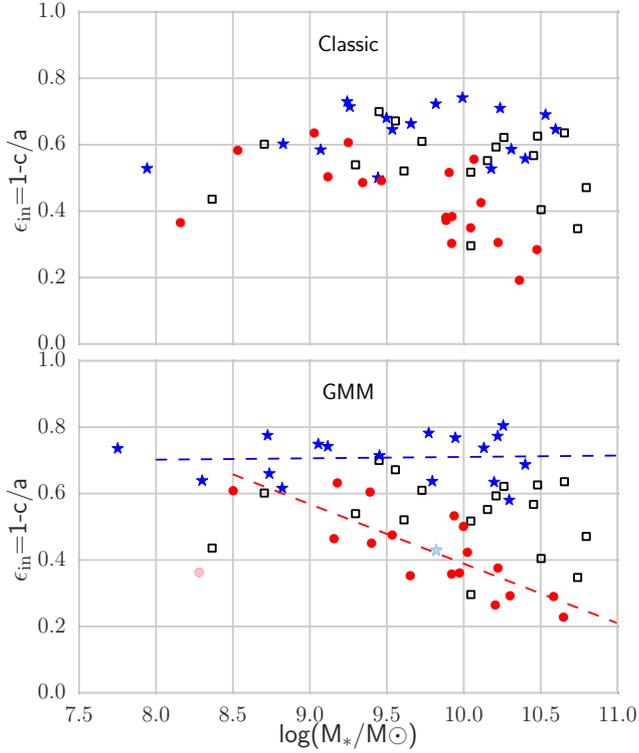}
 \caption{Intrinsic ellipticity, $\epsilon_{\rm in}$, as a function of stellar mass. The simulated spheroids and discs 
 defined using the classic method (top) and GMM (bottom) are given as filled red circles and blue stars, respectively. 
 The empty black squares give the full galaxies. The dashed blue and red lines in the bottom panel are linear regressions 
 through the disc and spheroid data, respectively. The symbols in light blue and pink have been excluded when fitting.}
 \label{fig_intrinsic_shapes}
\end{figure}

The basic property of a disc is its flat shape. 
In other words, discs have one axis that is much shorter than the other two and the large axes are roughly the same size. 
When a gas cloud is rotationally supported, it will flatten to the thickness supported by thermal pressure 
and maintain a radial size according to how fast it is spinning \citep{Fall:1980, Mo:1998}.

The intrinsic shape of galaxies can be quantified using the inertia tensor. 
Following \citet{GonzalezGarcia:2005}, the inertia tensor for a group of particles is defined as:
\begin{equation}
 I_{ij} = \sum_k m_{(k)}(\delta_{ij}r_{(k)}^2-x_{i(k)}x_{j(k)}),
 \label{eq6}
\end{equation}
where $k$ runs over all particles and $i$($j$) over the Cartesian coordinates. 
Diagonalizing this matrix produces the eigenvalues E$_1$$\leq$E$_2$$\leq$E$_3$, that can 
be used to compute the lengths of the three semiaxes $a\eqslantgtr b\eqslantgtr c$ as follows:

\begin{equation}
  \label{eq7}
  \begin{aligned}
    a^2+b^2+c^2 &= 5(E_1+E_2+E_3)/2\\
    a^2/b^2 &= (E_3+E_2-E_1)/(E_1+E_3-E_2)\\
    a^2/c^2 &= (E_3+E_2-E_1)/(E_1+E_2-E_3)
  \end{aligned}
\end{equation}

In this framework, the 3D shape of the discs and spheroids can be quantified by the deviation from symmetry, or the so-called 
intrinsic ellipticity parameter, $\epsilon_{\rm in} \equiv 1-c/a$. 
Flat discs have high ellipticity values, while rounder spheroids have values closer to 0. 

Fig.~\ref{fig_intrinsic_shapes} shows the intrinsic ellipticities of the galaxies (black squares) and their disc 
(blue stars) and spheroid (red circles) as a function of stellar mass. 
The GMM decomposition (bottom panel) reveals separate correlations with mass for the discs and spheroids.
The disc intrinsic ellipticity remains constant, $\epsilon_{\rm in} \sim 0.7$ as a function of mass. 
In contrast, the shape of spheroids changes significantly with mass. 
Spheroids in high mass galaxies are round, but spheroids in lower mass galaxies become flatter with a well defined slope 
of $\Delta \epsilon_{\rm in} \sim 0.18$ per decade of stellar mass (red dashed line). 
While this relation makes it look like spheroids will share the same shape as discs at low masses, there is little overlap in the GMM sample.

In the ''classic`` decomposition there is significant overlap between the regions occupied by the spheroids and discs.
Also, the correlation intrinsic ellipticity -- mass for the spheroid has more scatter.

Neither decomposition finds perfectly round spheroids. 
While spheroids are expected to be mildly tri-axial, the relatively high flatness of the lower mass spheroids indicates they may contain disc, namely a thick disc.
Our simple two component decomposition explains the spheroid flatness. 
The spheroid includes not only a bulge, but potentially a bar, pseudobulge or/and thick disc.
This figure suggests that at the low mass end, 
there is little differentiation between spheroids and discs, both having the appearance of thick discs.

\begin{figure}
 \includegraphics[width=0.5\textwidth]{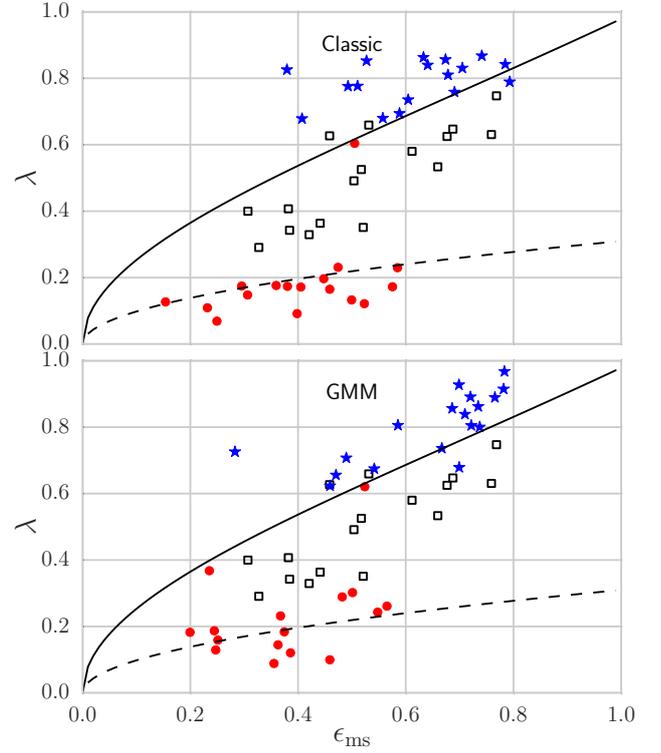}
 \caption{The spin parameter, $\lambda$ as a function of \emph{measured} ellipticity, $\epsilon_{\rm ms}$, within 0.1r$_{\rm vir}$. 
 The symbol and color codes are the same as in Fig.~\ref{fig_intrinsic_shapes}. 
 The solid black curve gives the locus of edge-on oblate spheroids flattened by rotation \citep{Illingworth:1977, Binney:1978, Cappellari:2007}. 
 The dashed black curve ($\lambda = 0.31\sqrt{\epsilon}$) separates slow (below) from fast (above) rotators.}
 \label{fig_rotsup_ellipticity}
\end{figure}

Fig.~\ref{fig_rotsup_ellipticity} relates the shapes to their kinematics. 
To allow easier comparison with observations, 
we derive properties as they are for IFU spectral observations. 
\citet{Emsellem:2004} found that $\lambda$ is a better indicator of rotational support than v/$\sigma$.
They defined $\lambda$ as:

\begin{equation}
\lambda \equiv \frac{\Sigma_i m_iR_i|v_{los,i}|}{\Sigma_i m_iR_i \sqrt{v_{los,i}^2+\sigma_{los,i}^2}}
\label{eq_lambda}
\end{equation}

In Equation~\ref{eq_lambda}, $i$ runs over the spaxels within an aperture of 0.1r$_{\rm vir}$, 
and m$_{\rm i}$ is the spaxel mass, R$_{\rm i}$ the distance to the centre, v$_{\rm los,i}$ the line of sight velocity and $\sigma_{\rm los,i}$ line of sight velocity dispersion.
The \emph{measured} ellipticity, $\epsilon_{\rm ms}$ is computed from the 2D mass distribution in the same 0.1r$_{\rm vir}$ aperture:

\begin{equation}
\epsilon_{ms} \equiv 1- \sqrt{\frac{\Sigma_i m_iy_i^2}{\Sigma_i m_ix_i^2}},
\label{eq_shape}
\end{equation} 
where y$_{\rm i}$ and x$_{\rm i}$ are the positions of the spaxel $i$ along the minor and major axes, respectively.
Both $\lambda$ and $\epsilon_{\rm ms}$ are computed from edge-on maps (see Appendix for more details).

In Fig.~\ref{fig_rotsup_ellipticity}, the solid black curve gives the theoretical prediction for 
oblate spheroids flattened by rotation \citep{Illingworth:1977, Binney:1978, Cappellari:2007}, while the dashed curve separates fast and slow rotators.
In this plane, the simulated galaxies (empty black squares) follow a curve offset to lower spins 
than the theoretical prediction. 
Both kinematic decompositions clearly separate discs from spheroids (high vs low spins and ellipticities). 
In both cases, discs appear above the black curve and spheroids are below.
Classic spheroids show less rotation (by construction), occupying the region of slow rotators (below the dashed black curve). 
The GMM discs show a correlation between shape and spin, while the classic ones seem 
to have almost constant $\lambda$, irrespective of shape. 
Also, the GMM discs reach higher $\lambda$ values, all the way up to the maximum, 1.  
Since the classic discs include more pollution from spheroids, their maximum $\lambda$ is 0.8.
Neither decomposition finds extremely flattened systems ($\epsilon_{\rm ms}>0.8$).

In observations, the spin and shape parameters are typically measured within the half light radius aperture. 
Given that we are interested in intrinsic properties across our simulation sample (all parameters are mass and not light weighted), 
we computed the spin and shape parameters in apertures which include the full galaxies (0.1r$_{\rm vir}$).
In aperture this large, we find the \emph{intrinsic} and \emph{measured} ellipticities to follow an approximately one-to-one correlation.
In smaller apertures (half mass radii for example) the \emph{measured} ellipticities are systematically smaller than the \emph{intrinsic} ones. 

\subsection{Ages}

\begin{figure}
 \includegraphics[width=0.5\textwidth]{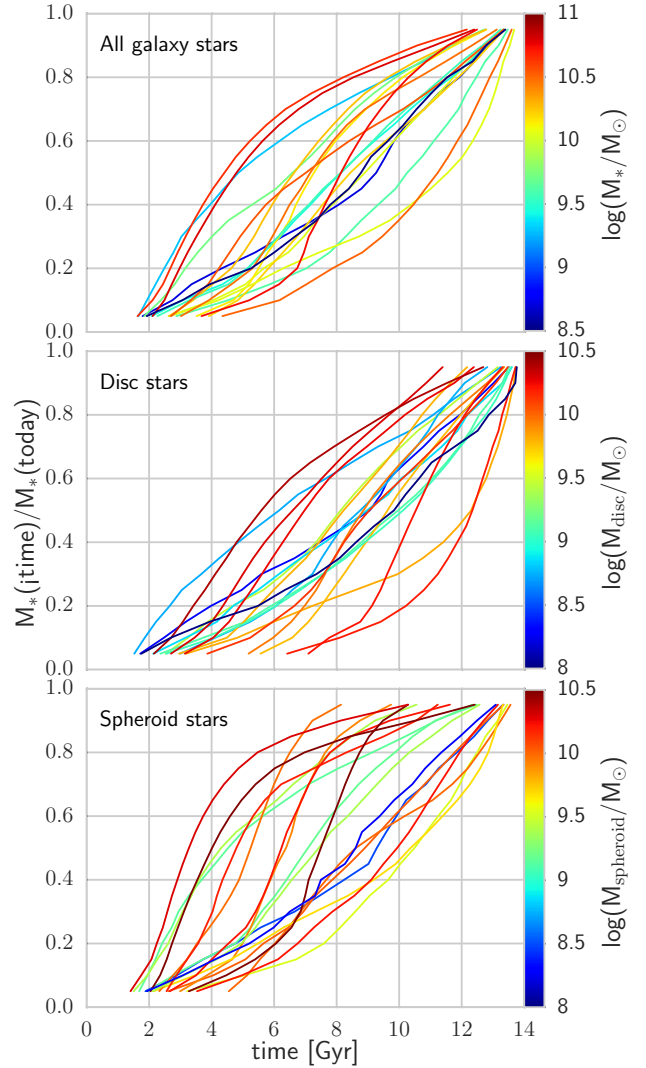}
 \caption{Cumulative stellar mass assembly tracks for the full galaxies (top), 
 their GMM discs (centre) and GMM spheroids (bottom), color coded by the corresponding stellar mass at z=0.}
 \label{fig:cumage}
\end{figure}

Determining the age of stars in discs and spheroids can constrain their formation histories.
Fig.~\ref{fig:cumage} shows cumulative star formation histories (SFHs) for each galaxy (top), 
GMM disc (middle) and spheroid (bottom).  The SFHs are
colored and normalized according to their final mass at $z=0$. 
The normalization allows for comparison of galaxies of different masses.
  
There is considerable variety among the SFHs of the full galaxies (top panel)
and no clear trend with final stellar mass. 
Some of the more massive (red curves) galaxies show elevated early star formation, 
while others show vigorous star formation at late epochs. 
The two lowest mass galaxies have almost constant specific star formation rates (sSFR) over 
long periods of time: $\sim$ 0.06 Gyr$^{\rm -1}$ for the first half of the Universe age
and $\sim$ 0.1 Gyr$^{\rm -1}$ for the second one. 
A more detailed look at SFHs shows that star formation is bursty on 100 Myr time-scales. 

A look at the component SFHs shows that spheroids of the massive galaxies rise sharply prior to 7 Gyr. 
About half of the spheroids show a constant sSFR of $\sim$ 0.1 Gyr$^{\rm -1}$. 
There are basically no spheroids with rates smaller than this value. 

The discs generally grow their mass slower than the spheroids, 
though two of the most massive objects appear to grow only in the last $\sim$4 Gyrs.
The most extreme case forms 80\% of its stars in the last 3 Gyr.
Most of the lower mass discs, however, grow their mass steadily over a long time.

\begin{figure}
 \includegraphics[width=0.5\textwidth]{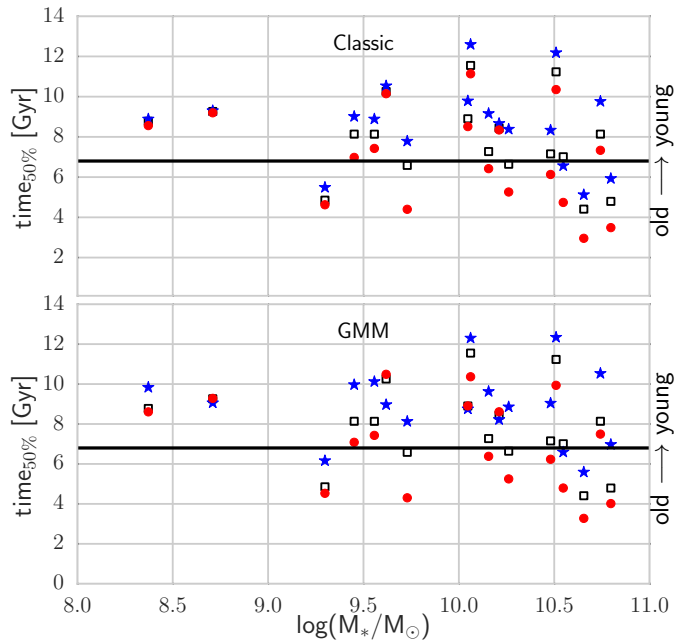}
 \caption{Half mass formation time as a function of total stellar mass for the simulated spheroids (filled red circles) and discs (blue stars),
 defined using the classic method (top) and GMM (bottom).
 The horizontal black line gives the half age of the Universe (6.8 Gyrs).}
 \label{fig:t50}
\end{figure}

In an attempt to boil complicated star formation histories down to one number, 
Fig.~\ref{fig:t50} shows the median age of stars as a function of the 
total stellar mass for the entire galaxy. 
We use the total galaxy mass so that it is possible to compare 
the ages of the discs and the spheroids of specific galaxies. 
At M$_{\rm *} > 10^{\rm 10} \msun$, the spheroids are consistently older than the disc by several Gyr. 
At M$_{\rm *} < 10^{\rm 10} \msun$, the ages of components are typically less separate.
At these lower masses, the ages converge down to a line slightly 
younger than half the age of the Universe, 6.8 Gyr. 
A median age around half the age of the Universe is in line with 
the interpretation of Fig.~\ref{fig:cumage} that 
lower mass galaxies steadily build up their stellar masses. 
Another name for more massive galaxies forming their stars 
earlier than low mass galaxies is ``downsizing''.

Fig.~\ref{fig:t50} also shows another subtle difference between the two decomposition methods.
The classic method always finds discs having 
younger stellar populations than their corresponding spheroids. 
In GMM, four of the eighteen galaxies have younger spheroids than discs. 

It is surprising that spheroids should be younger.  It is conventional wisdom
that stars always form in thin discs.  Younger spheroids mean that
stars could form with non-negligible vertical velocity dispersion.  Such
young bulges only form in low mass galaxies where stellar feedback
is able to drive turbulence and trigger star formation out of the disc plane.
It is interesting that the classic decomposition that assumes spheroids have no net rotation, 
finds older bulges as one would expect from galactic components that have undergone
significant merging.

\begin{figure}
 \includegraphics[width=0.5\textwidth]{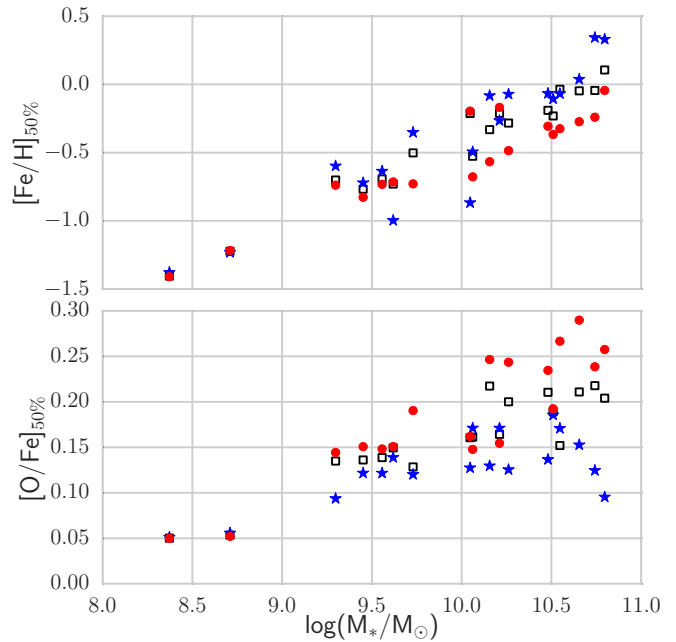}
 \caption{Median [Fe/H] (top) and median [O/Fe] (bottom) as a function of total stellar mass for the GMM spheroids (filled red circles) and discs (blue stars).
 The empty black squares give the full galaxies.}
 \label{fig:mzr}
\end{figure}

\subsection{Chemical abundances}
Observationally determining stellar ages of individual stars is often complicated. 
Ages require resolving or taking spectra of 
individual stars.  In some cases, age dating might only provide rough ages for entire populations. 
An easier thing to observe is the metal abundance of galaxies. 
Metals are the result of star formation, so the chemical abundance of galaxies 
gives a rough idea of when various components of the galaxy formed. 
In this way, it can give a rough observational estimate of the formation history of a galaxy. 

One simple metric is the ratio between $\alpha$ elements like 
oxygen, magnesium or calcium and iron. 
All these elements are produced in core collapse supernovae (SNII) 
of massive stars that happen soon after the birth of stellar populations. 
Over a longer time-scale, type Ia supernovae (SNIa) enrich galaxies with only iron peak elements. 
Thus, later enrichment from SNIa lowers the ratio of $\alpha$ elements to 
iron and any stars that form thereafter will have the chemical signature of a lower 
(solar) $\alpha$-to-iron ratio. 
So, galactic components that are dominated by stellar populations 
that form quickly and do not have much subsequent star formation will be $\alpha$-rich, 
while populations that formed over a long time will be $\alpha$-poor.  
\citet{Stinson:2013b} found that [$\alpha$/Fe] is a better chronometer than [Fe/H].

The upper panel in Fig.~\ref{fig:mzr} shows the median metallicity, [Fe/H], of the simulated galaxy 
components as a function of their stellar mass. 
[Fe/H] increases steadily with stellar mass.
For M$_{\rm *}>$10$^{\rm 9}$M$_{\odot}$, at fixed stellar mass, discs are systematically more metal rich than spheroids,
which generally form their stars earlier.
We show only the GMM classification because the differences from the classic one are only minor. 
The [Fe/H] of the discs and spheroids are slightly better separated for the GMM classification than for the classic one.

The lower panel Fig.~\ref{fig:mzr} shows the median [O/Fe] for each component as a function of stellar mass. 
The [O/Fe] relation with stellar mass is different than [Fe/H]. 
Instead of [O/Fe] steadily increasing, [O/Fe] $\sim0.05$ for M$_{\rm *}<10^{\rm 9} \msun$. 
Most of the higher mass total galaxies and components have median [O/Fe] $\sim 0.15$.
The only exception are a few high-mass spheroid components that have [O/Fe] $\sim 0.25$. 
These spheroids have experienced intense early periods of star formation.
The higher mass galaxies are also the ones that are more disc dominated. 
This directly translates into continuous levels of star formation throughout the disc history, 
which increase the oxygen abundance. 
However, the initial high levels of star formation ensure that 
many SNIa enrich the galaxy with iron, decreasing [O/Fe].
We show only the GMM classification in this figure since the classic one gives very similar results. 

As with the other properties examined, the difference becomes less in lower mass galaxies.

\subsection{Surface density profiles}
\label{sdp}
In general, the 2D and 1D light profiles of galaxies are well described by S\'{e}rsic functions. 
One common assumption about observed galaxies is that a galaxy with a S\'ersic index $n\sim1$, 
which is close to exponential, is characterized as a disc. 
Fig.~\ref{fig_mass_dt} showed that most of the simulated
galaxies have pure exponential light profiles, so 
it is interesting to look at the actual mass profiles. 

\begin{figure*}
 \includegraphics[width=0.98\textwidth]{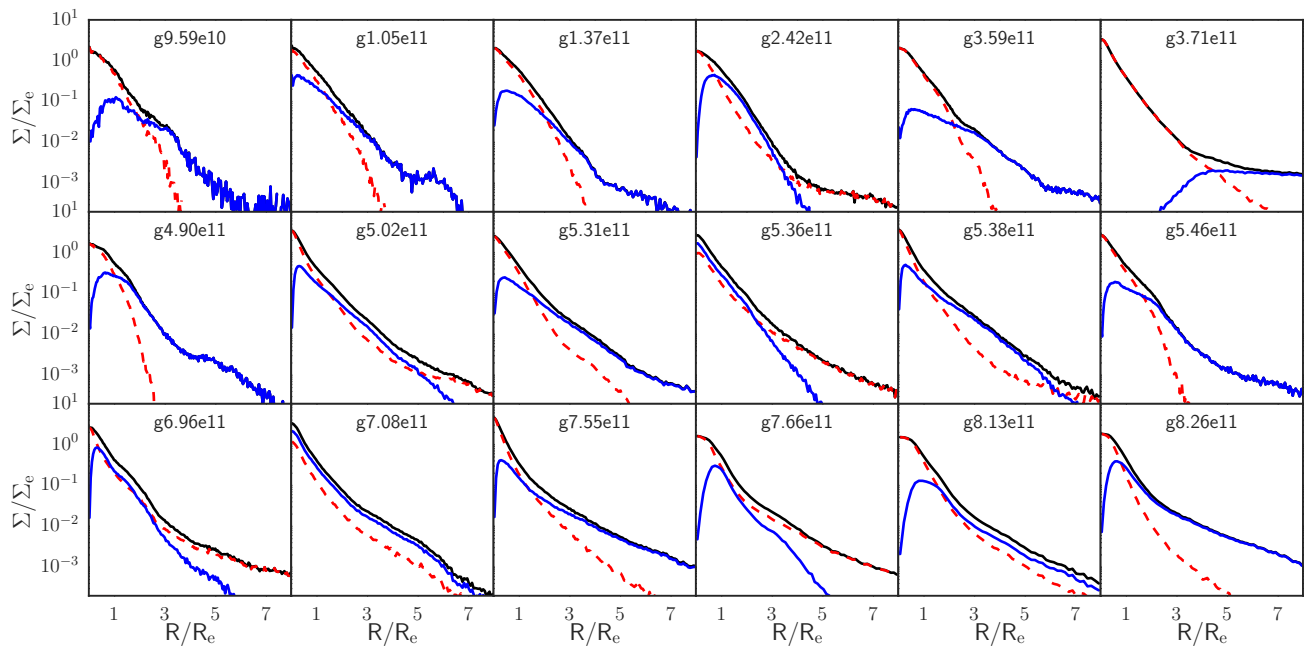}
 \caption{The face-on mass surface density profiles of the simulated galaxy sample (solid black curves), 
 and of their GMM discs (solid blue curves) and GMM spheroids (dashed red curves).
 The profiles have been normalized to the effective mass surface density of the full galaxies ($\Sigma_{\rm e}$) and the radii to the effective radius (R$_{\rm e}$) of the full galaxies.}
 \label{fig:profiles}
\end{figure*}

Fig.~\ref{fig:profiles} shows all eighteen surface mass density profiles 
for GMM discs (solid blue), GMM spheroids (dashed red) and full galaxies (solid black). 
The profiles have been normalized to $\Sigma_{\rm e}$ of the full galaxies and 
given as a function of R/R$_{\rm e}$. R$_{\rm e}$ is the half mass radius of the full galaxies viewed face-on. 
About 2/3 of the less massive galaxies (black curves on the top and centre rows)
are exponential to at least 5R$_{\rm e}$.
Based on mass profiles only, one would be tempted to classify these galaxies as pure discs. 
However, our kinematic decomposition showed two distinct components. 
At higher masses, central concentrations start to appear.  

In the same figure, the kinematic discs (blue) and spheroids (red) show a wide range of behaviors. 
For example, g1.05e11, g5.31e11 or g7.55e11 show discs and spheroids as one would expect from a simple two component $\chi^{\rm 2}$ fit. 
Their spheroids dominate the central regions, while their discs have exponential profiles and dominate the external parts. 
Other galaxies, like g2.42e11 and g5.02e11 have exponential discs, but spheroids that dominate both 
the innermost and the outermost regions (probably their bulges and stellar haloes have different origins). 
g7.08e11 also shows a rather peculiar behavior, being a clearly disc dominated galaxy with a spheroid profile a 
scaled version of the disc one. The fact that this galaxy is disc dominated is also clear from both the {\sc sunrise} image 
(Fig.~\ref{fig:sunrise}) as well as from the circularity distribution (Fig.~\ref{figure11}).

In any case, two common features show up in almost all the galaxies of the sample.
The spheroids do not have strongly central peaked profiles, but look rather smooth and close to exponentials. 
On the other hand, discs  can be described by exponentials on extended ranges and in most cases do not extend to the innermost parts. 
Most of the disc profiles show central dips, simply because the GMM decomposition associates very few of the inner stellar particles 
with them. However, observationally discs are assumed to extend all the way to the galaxy centre. 

The most surprising aspect of all of these simulations is how close 
the kinematically selected spheroids are to exponential density profiles. 
Observationally, some dwarf galaxies are not thin discs, but have exponential profiles \citep{Graham:2003}.
It seems that an exponential density profile is an insufficient discriminant between discs and non-discs. 
Fig.~\ref{fig:profiles} points out that, in the absence of information on the 
stellar dynamics, surface density profiles are not sufficient to classify galaxies 
as kinematic discs with purely circular orbits.

\subsection{S\'{e}rsic indices}
When fitting profiles, various problems can occur 
due to the large number of parameters, the binning, or the fitting range. 
These issues can be avoided by using non-parametric defined quantities, 
like the radius that encloses a fixed fraction of the total mass. 
Therefore, rather than fitting the S\'{e}rsic profile directly,
we use the corresponding mass enclosed inside a given radius $R$: 

\begin{equation}
\begin{aligned}
 M(<R) &= 2\pi\int_0^R \Sigma(x)xdx\\ 
 ~~~~~ &= 2\pi R_e^2\Sigma_e\frac{n e^{b_n}}{b_n^{2n}}\gamma(2n,b_n(R/R_e)^{1/n})
\end{aligned}
\label{eq2}
\end{equation}
In Equation~\ref{eq2}, $\Sigma_{\rm e}$ is the surface mass density at half mass (or effective) radius, R$_{\rm e}$,
$\gamma$ is the lower incomplete Gamma function and b$_n$ is approximated by $b_n \simeq 1.9992n - 0.3271$. 
$\Sigma(R)$ is the S\'{e}rsic function $\Sigma_{\rm e} exp(-b_{\rm n}((R/R_{\rm e})^{\rm 1/n}-1))$.
Using Equation \ref{eq2} for two different fractions of the total mass, $f_{\rm 1}$ and $f_{\rm 2}$ (for example 90\% and 50\%),
results the following equation for the S\'{e}rsic index that can be solved numerically:

\begin{equation}
 \frac{\gamma(2n,b_n(R_{f_1}/R_e)^{1/n})}{f_1}-\frac{\gamma(2n,b_n(R_{f_2}/R_e)^{1/n})}{f_2}=0
 \label{eq3}
\end{equation}

The values of the S\'{e}rsic index in Table \ref{table2} 
are computed in this manner using the radii enclosing 90 and 50\% of the mass, respectively. 

\begin{figure}
 \includegraphics[width=0.5\textwidth]{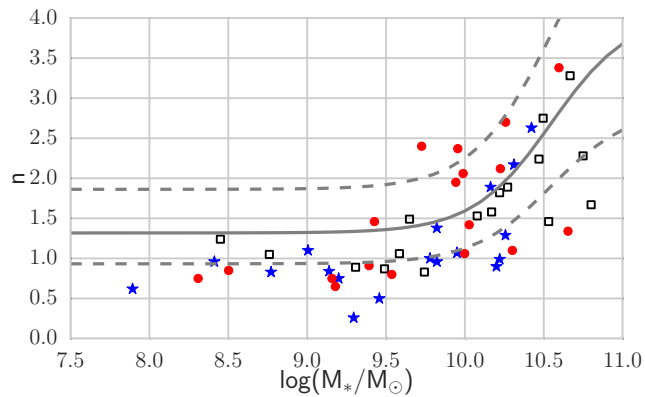}
 \caption{The S\'{e}rsic index derived using Equation \ref{eq3} as a function of stellar mass for the GMM disc (blue stars),
 GMM spheroid (filled red circles) and all (filled black squares) galaxy stars.
 The solid grey curve gives the variation of n with stellar mass as parametrized by \citet{Dutton:2009} using  
 observations of SDSS late type galaxies \citep{Blanton:2005}, while the dashed curves give the corresponding 1$\sigma$.}
 \label{fig_mass_n} 
\end{figure}

Fig.~\ref{fig_mass_n} gives the S\'{e}rsic index, n, derived using Equation \ref{eq3} 
as a function of stellar mass for all the simulated galaxies in the sample. 
The solid curve comes from \citet{Dutton:2009}'s parameterization of the n-M$_{\rm *}$ correlation
for blue galaxies. \citet{Dutton:2009} split SDSS observations of galaxies 
\citep{Blanton:2005} into red and blue samples according to their colors. 
He fit the median n as a function of stellar mass.
The low mass median blue galaxies in the SDSS sample have S\'{e}rsic indices close to 1.5.
That means they are nearly exponential, but the median galaxy is best-fitted with a slight bulge.
While there are fewer blue galaxies above M$_{\rm *}=2.5\times10^{\rm 10} \msun$ than below,
the S\'{e}rsic indices increase at this transition mass.  There are more red galaxies at these
higher mass that also have S\'{e}rsic indices near 4.  The combined SDSS sample including both
red and blue galaxies shows a similar transition to the discs, but at slightly lower mass.

We choose to compare our simulations with the blue sample since all of our galaxies are
forming stars at the $z=0$.  Also, our sample was selected to include the most discy (flat) galaxies.  
Additionally, NIHAO contains only isolated galaxies, which have a lower likelihood of being red
than galaxies that form in a denser environment.

The simulated sample does not include enough galaxies to make a rigorous quantitative
comparison with the observed S\'{e}rsic indices. 
We base our comparison on the black squares that show the S\'{e}rsic for the entire galaxy. 
Qualitatively, the simulations follow the shape of the observed relation. 
The simulated galaxies at M$_{\rm *}<10^{\rm 10} \msun$ have S\'{e}rsic indices around 1. 
Above M$_{\rm *}>10^{\rm 10} \msun$, the S\'{e}rsic indices generally rise following a similar trend to the observations.

As opposed to mass profiles, light profiles can be effected by dust, 
which makes the centre of galaxies appear to have less of a bulge and hence lower the S\'{e}rsic indices.

There is a slight hint that the simulations have lower S\'{e}rsic indices than observations.
Low S\'{e}rsic indices might indicate that the stellar feedback could be too strong in the {\sc nihao} sample. 
However, a definite statement would require a larger sample of galaxies and is beyond the scope of this paper. 
It is also interesting that deep photometric imaging have revealed low mass ($10^8 \msun$) galaxies
with S\'{e}rsic indices below 1 \citep[see observations by][]{vanDokkum:2015, Koda:2015}.

The individual kinematic components generally follow the entire galaxy S\'{e}rsic indices.
From the lowest masses, this is particularly true.  The surface density profiles of material
traveling along circular orbits, the disc, has a similar spatial distribution to material that
has higher velocity dispersion (spheroid). 
Above M$_{\rm *}>10^{\rm 9.5} \msun$, the components start to show independent distributions. 
The discrepancy in S\'{e}rsic index then increases with mass.
For the most part, the spheroidal component has a higher S\'{e}rsic index, 
though there are a couple cases where the disc has higher $n$.  

\section{Summary and conclusions}
\label{conclusions}
We present a new kinematic decomposition technique and use it to
decompose a subsample of the {\sc nihao} suite of zoom-in 
cosmological simulations \citep{Wang:2015} at $z=0$ into discs and spheroids.
The shapes, kinematics, ages, and abundances of the discs are compared with those of the spheroids.
Additionally, the structural parameters of the {\sc nihao} galaxies are compared with observations. 

The new decomposition technique minimizes ambiguous particle classification across a large sample of
galaxies with a variety of morphologies. It thus enables future studies of the 
formation and evolution of galaxy components. 

There are three main findings of the study:
\begin{itemize}
\item While the properties of discs and spheroids differ as expected in high mass galaxies, 
they merge below 10$^{\rm 9.5}\msun$. Such low mass galaxies are defined
by a structure that is somewhere in between thin discs and spheroids, something resembling a thick disc.
\item Many of the galaxies in our sample exhibit pure exponential surface brightness profiles,
a signature typically associated with pure disc galaxies.
However, the kinematic decomposition finds thin disc fractions less than 50\% in many galaxies
showing that surface brightness profiles are not an optimal indicator of galaxy kinematics.
\item The kinematic properties of the GMM discs agree well with observations of disc dominated systems.
\end{itemize}

\subsection{Decomposition technique}
The method builds on previous work by DM12,
but employs Gaussian Mixture Models to find groups in a 3D parameter space ($j_{\rm z}/j_{\rm c}$, $e$, $j_{\rm p}/j_{\rm c}$) rather than \textit{k-means}.

With numerical simulations, it is possible to compare the photometric structure with 
the kinematic structure since the simulation outputs all six position and velocity variables.
Most of previous kinematic decompositions of simulations have focused on one variable, the angular momentum of stars, and assume the spheroid has no net rotation.

When our decomposition was compared with the ``classic'' one, we found some subtle differences.
The biggest difference is that the GMM method makes a clear selection of disc and spheroid particles. 
Thus, the GMM discs show more coherent rotation than the classic method and a flatter shape.
Somewhat surprisingly, spheroids of low mass galaxies also show some rotation and flatter shape using the GMM decomposition.

The classic decomposition method always finds spheroids are older than discs.
In a few cases, the GMM method finds spheroids with median ages \emph{younger} than their corresponding discs. 
How this happens in the simulations will require a follow-up study,
but could have significant implications for the galaxy formation paradigm.  
It could provide evidence that stars do not always form in thin discs. 

\subsection{Disc--spheroid comparison}
Kinematically defined discs are thinner than spheroids, typically younger, more metal rich, have lower [$\alpha$/Fe], lower S\'{e}rsic indices 
and higher spin parameters.

The properties of discs and spheroids showed significant trends as a function of galaxy mass. 
At low masses, the distinction between discs and spheroids disappeared in their velocity dispersion, 
age, chemical enrichment and S\'{e}rsic index.
The properties start to differ only above M$_{\rm *} \gtrsim 10^{\rm 9.5} \msun$.
We note that \citet{Simons:2015} observe a similar break in 
galactic kinematic properties at exactly this same mass. 

The $j_{\rm z}/j_{\rm c}$ distribution functions of the lowest mass galaxies are in agreement with a unique slowly rotating spheroid component,  
indicating that demanding that two components exist may be inappropriate. 
However, the velocity maps show signatures of coherent rotation even in these low mass systems. It is, thus likely that they have some sort of disc as component.

The low mass galaxies all show nearly exponential stellar surface density profiles
despite including a large fraction of stars with high velocity dispersions.  
When broken down into disc and spheroid components, 
\emph{both} components exhibit exponential profiles.
Thus, our results show that there is \emph{no direct relationship between exponential profiles and circularly rotating thin discs}.

\subsection{Comparisons with observations}
The specific angular momenta, velocity dispersions and S\'{e}rsic indices of the simulated galaxies are compared with the analogue properties of observed galaxies. 
While the comparisons are non-trivial because of selection effects and limited sample sizes, the simulated galaxies show similar
properties as the observed galaxies in all cases. 

The simulated galaxies fall on top of observed ones in the stellar mass - specific angular momentum plane. 
When broken down into kinematic components, both simulated discs and spheroids show a shallower correlation between specific angular momentum and mass 
than the theoretical expectations for pure spirals and elliptical galaxies, but consistent with observations.

The S\'{e}rsic indices of the simulated galaxies vary with stellar mass as do late type SDSS galaxies for M$_{\rm *}>$10$^{\rm 9.5}$M$_\odot$.
For smaller stellar masses $n\sim1$. 

Previous studies have shown that simulations using stellar feedback to limit star formation produce galaxies that are too thick.
{\sc Nihao} does not clarify that result.  
The simulations, including the kinematically cold thin disc component, are three times thicker
than the thin discs in the sample of edge-on Spitzer galaxies presented by \citet{Comeron:2011}.
Despite this, the velocity dispersions of the thin disc components match the vertical velocity
dispersions from the \citet{Martinsson:2013}'s kinematically cold IFU \emph{DiskMassSurvey}.  
The galaxies and their spheroid components are about the same thickness as \citet{Comeron:2011}'s observed thick discs.  
Perhaps unsurprisingly, the total galaxies have twice the velocity dispersion of the kinematically cold DMS sample.

Stellar feedback forms galaxies with thick discs that are typically more massive than observed.
Using the GMM decomposition technique, it is possible to classify stars into distinct kinematically cold and hot components.  
Following up on the formation and evolution of these components will provide insights into how 
to improve galaxy formation simulations and begin to provide detailed answers to how galaxies really form.

\section*{Acknowledgments}
The simulation codes used in this project were developed by a large team led by Tom Quinn and James Wadsley. Without their contributions, this project would have been impossible.
The authors are grateful to Julianne Dalcanton, 
Hans-Walter Rix, Marie Martig, Remco van den Bosch, 
Ling Zhu, Peter Yoachim, Glenn van de Ven and Rosa Dom\'{\i}nguez Tenreiro
for useful conversations and comments.
The simulations were performed on the \textsc{theo} and \textsc{hydra} clusters of  the
Max-Planck-Institut f\"ur Astronomie at the Rechenzentrum in Garching. 
We greatly appreciate the contributions of these computing allocations.
AVM, AAD and GSS acknowledge support from the  Sonderforschungsbereich SFB 881 ``The Milky Way System''
(subproject A1) of the German Research Foundation (DFG).   
GSS acknowledges funding from the European Research Council under the European Union's Seventh Framework Programme (FP 7) ERC Grant Agreement n. [321035].
XK is supported by the NSFC (No.11333008) and the strategic priority research program of CAS (No. XDB09000000).

\bibliographystyle{mnras}
\bibliography{AObreja}

\section*{Appendix}
\label{appendix_section}
\appendix
\renewcommand{\thefigure}{A\arabic{figure}}
\renewcommand{\thetable}{A\arabic{table}}
\renewcommand{\theequation}{A\arabic{equation}}
\setcounter{figure}{0}
\setcounter{table}{0}
\setcounter{equation}{0}

Here we describe the manner in which we compute line of sight velocities and velocity dispersions maps for our simulations.
We also discuss the way in which we derive simulation analogues to observables from these mass weighted maps. 

In observations, IFU data cover a limited field of view by a 2D regular grid (Cartesian or hexagonal) 
with varying signal-to-noise (S/N) from pixel to pixel, 
without considering the S/N dependency on the wavelength range. 
Therefore, quantities computed from these maps have to take into account the varying S/N. 
A more general method however, is to rebin the 2D regular grid as to ensure an approximately uniform S/N and 
then generate a Voronoi mesh from the centroids of the rebinned grid. 
This approach has been implemented by \citet{Cappellari:2003} for the SAURON Project \citep{Bacon:2001}. 
The same method has been used for other IFU surveys like ATLAS3D \citep{Cappellari:2011}, 
as well as to post-process simulated galaxies \citep[e.g.][]{Naab:2014}. 

In this study, we are concerned with intrinsic properties of the simulations. 
Binning and rebinning the SPH particle data would allow for a closer comparison with observations.
However, it would also add uncertainties. 
As such, we chose to do the Voronoi tessellation directly on the stellar particles.
In this way, each of the 2D spaxels has approximately the same number of SPH particles. 
This is equivalent to the constant S/N spaxels in real IFU data. 

In observations, the different moments of the velocity fields are computed by fitting stellar absorption lines of each spaxel separately. 
A common choice for the fitting function is the Gauss-Hermite \citep{Bender:1994}.
The equivalent in simulations is to fit the distribution of particles' line-of-sight velocities in each spaxels. 
However, instead of assuming a functional form for the velocity PDFs (which might or might not be appropriate), 
we choose to use nonparametric statistics.
Therefore, we compute the line-of-sight velocity (v$_{\rm los}$) and line-of-sight velocity dispersion ($\sigma_{\rm los}$) of each spaxel  
as the mass weighted mean and variance of the empirical distribution of particle velocities (v$_{\rm i}$):

\begin{equation}
  \begin{aligned}
    v_{los} &= \Sigma_i w_iv_i\\
    \sigma_{los}^2 &= \Sigma_i w_i(v_i - v_{los})^2\\
  \end{aligned}
\label{eq_appendix}  
\end{equation}

In Equation~\ref{eq_appendix}, $i$ runs over the particles in the spaxel and w$_{\rm i}$=m$_{\rm i}$/$\Sigma_{\rm i}m_{\rm i}$ is each particle's weight. 
The Voronoi tessellations for all the galaxies were constructed requiring 1024 elements. 
The tessellation produces approximately equal number of particles in each spaxels. This translates into some spaxels (mostly close to the border of the domain)
having a smaller number of particles. Therefore, when computing the statistics for each of them, we tag as ``bad spaxels'' those that contain less than 30 particles.
The tessellations and maps have been constructed for the galactic region, defined as a sphere with radius 0.1r$_{\rm vir}$.

\begin{figure*}
\includegraphics[width=0.75\textwidth]{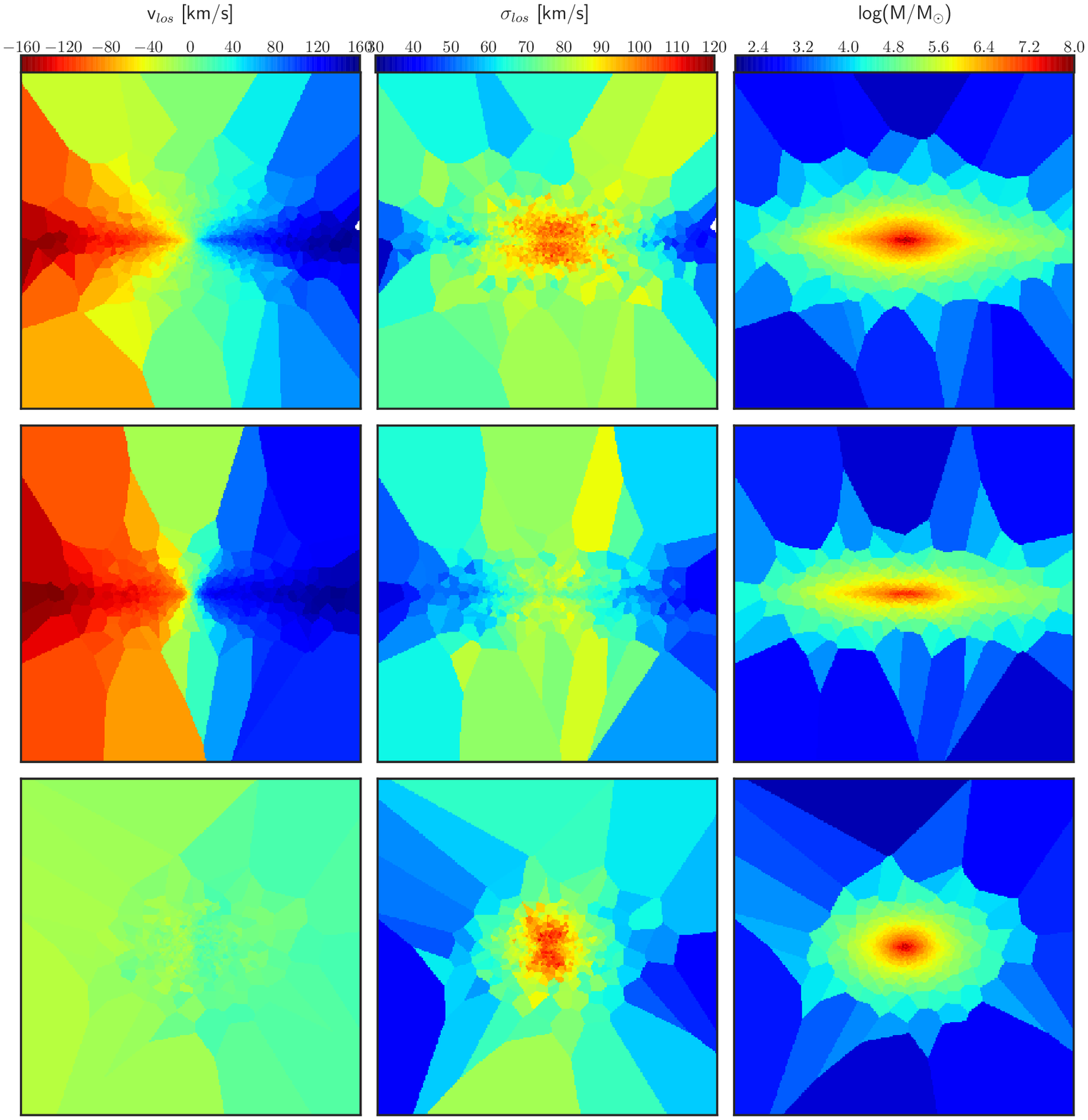}
\caption{Line of sight velocities (left column), line of sight velocity dispersions (central column)
and stellar mass distributions (right column) for one edge-on view of the simulated galaxy g7.55e11 considering all stars within 
0.1r$_{\rm vir}$ (top row), and only those (within the same region) classified as belonging to the GMM defined disc (central row) and 
spheroid (bottom row). The maps in each column share the same intensity scale, as given by the upper color bars. 
The panels are 42 kpc across.}
\label{fig_velocitymaps}
\end{figure*}

Fig.~\ref{fig_velocitymaps} shows an example of v$_{\rm los}$ (left column) and $\sigma_{\rm los}$ (central column) maps 
from an edge-on view of the simulated galaxy g7.55e11 (top row), and its GMM disc (central row) and GMM spheroid (bottom row). 
The right column of the figure gives the corresponding stellar mass distributions.
Each column in the figure shares a common color coding in order to facilitate the comparison between the full galaxy and its kinematic components. 

The upper left panel of the figure gives the v$_{\rm los}$ of the full galaxy. It shows unambiguously the signature of a rotating disc. 
The corresponding $\sigma_{\rm los}$ map (upper central panel) also shows a distinctive feature: 
two central and symmetric dispersions maxima above and below the equatorial plane. The dispersion along the major axis is almost half the one along the minor axis,
and decreases with the distance from the centre. In the stellar mass distribution panel (upper right), the full galaxy is clearly a highly flattened system. 

Looking at the two kinematic components, we see that the disc is mainly responsible for the rotation signature in the full galaxy.
The spheroid is responsible for the double velocity dispersion peak. 
The disc, however, also shows a double $\sigma$ peak but less pronounced 
($\sim$80 km~s$^{\rm -1}$ as compared to $\sim$120 km~s$^{\rm -1}$).
The equivalent goes for the spheroid, which also rotates, but much slower than the disc ($\sim$40 km~s$^{\rm -1}$ as compared to $\sim$160 km~s$^{\rm -1}$).
The disc and spheroid mass distributions in the right column complement the velocity information. 
The former is clearly a highly elongated structure, while the later is approximately spherical. 

These kind of maps can provide more than a qualitative idea of galaxy structure. 
A whole range of parameters that can be used to quantify the kinematic structure of galaxies
can be extracted from them \citep[see for example][]{Cappellari:2007}. 
In this study, we have analysed some of these parameters, namely: spin ($\lambda$), \emph{measured} ellipticity ($\epsilon_{\rm ms}$),
central vertical velocity dispersion (v$_{\rm z,0}$), and maximum rotational velocity (v$_{\rm max}$). 

\begin{figure*}
 \includegraphics[width=1.0\textwidth]{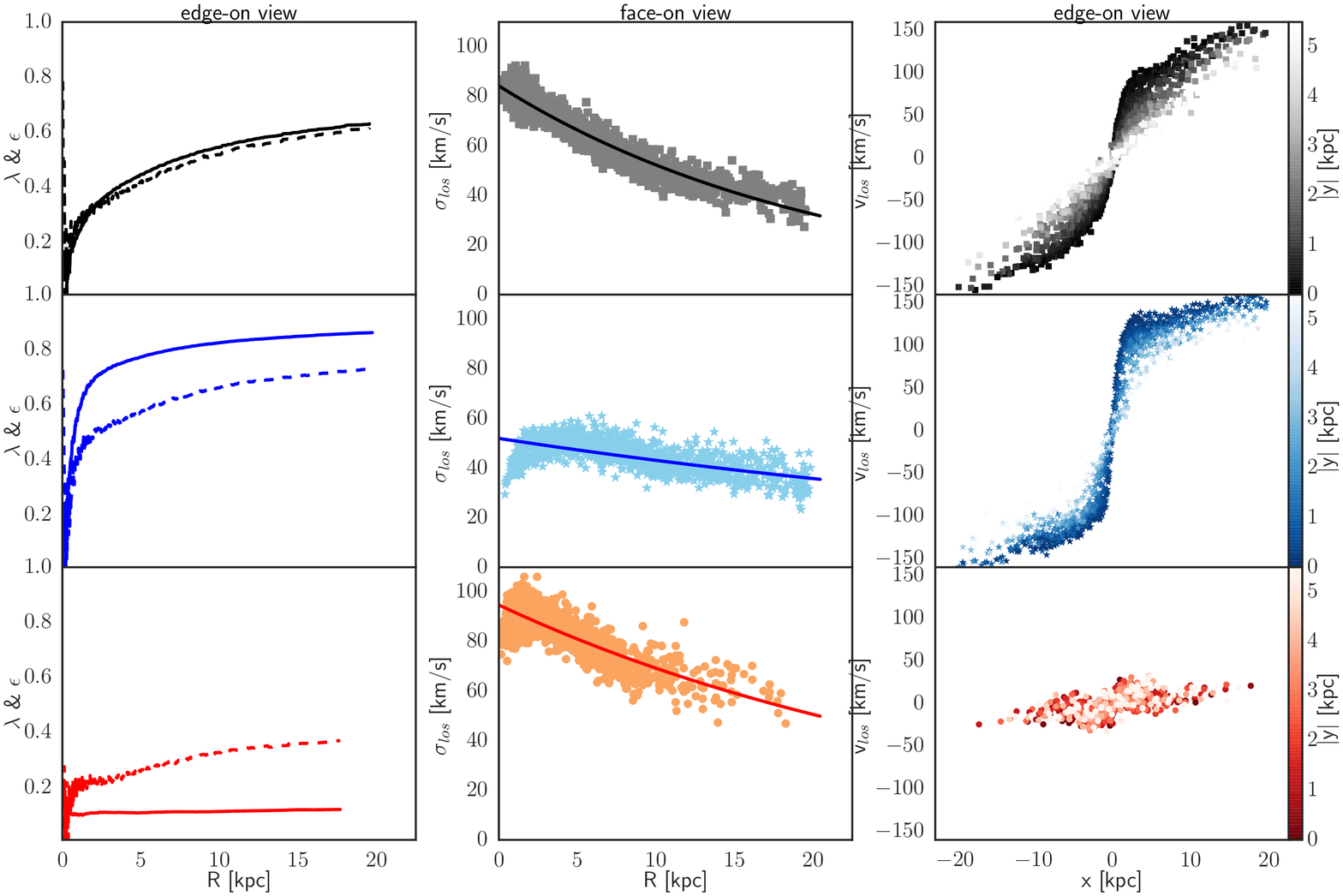}\\
\caption{\textbf{Left column}: radial profiles of the spin (solid curves) and measured ellipticity (dashed curves) 
from an edge-on view of the galaxy g7.55e11 (top), and its GMM defined disc (centre) and spheroid (bottom).   
\textbf{Central column}: radial dependence of the line of sight velocity dispersion from a face-on view
of the same galaxy (top) and its GMM disc and spheroid (central and bottom rows, respectively). 
The black, blue and red curves are the exponential fits to the grey, light blue and light red symbols, respectively. 
The central line of sight velocity dispersions, $\sigma_{\rm z,0}$ of Fig.~\ref{fig_velocity_sigma} 
are the central values from these exponential fits.
\textbf{Right column}: Line of sight velocities as functions of radial distance color coded by the distance from the 
equatorial plane for the same galaxy viewed edge-on and its two GMM components (top to bottom).}
\label{fig_faceon_and_edgeon_radialparam}
\end{figure*}

Fig.~\ref{fig_faceon_and_edgeon_radialparam} gives a graphical description of how the four parameters mentioned above 
are derived from the maps. 
As for the previous figure, the top, central and bottom rows give the full galaxy, the GMM disc and the GMM spheroid of g7.55e11. 
The spin and ellipticity parameters shown in the left column are optimally derived when galaxies are viewed edge-on. 
Same goes for the rotation curve (right column). 
The central vertical velocity dispersion, however, is optimally derived when the galaxy is viewed face-on (central column).

The spin and ellipticity (solid and dashed curves in the left column) have been computed with equations \ref{eq_lambda} and \ref{eq_shape}
in circular apertures, up to the maximum one of 0.1r$_{\rm vir}$. 
In observations, elliptical apertures are used instead. Since we weight the spaxels with the mass, 
the bias introduced by circular apertures when the maps show elongated features should be small. 
The parameter most sensible to the aperture choice is $\epsilon_{\rm ms}$. 
In this case, we have checked that the \emph{measured} ellipticity in an aperture of 0.1r$_{\rm vir}$
is actually a very good estimator for the \emph{intrinsic} 
ellipticity derived from the inertia tensor of the particles within the same radius. 
Therefore, we are confident that changing the aperture from circular to elliptical will not alter our conclusions. 

The central column of the figure shows the values of all spaxels $\sigma_{\rm los}$ as a function of radius, R. 
The solid curves in these panels are the exponential $\chi^{\rm 2}$ fits to the simulation data. 
Each spaxel contribution has been weighted by its mass. 
In this manner we derived the central vertical velocity dispersions that were used in Section~\ref{dms_comparison}.
In order to obtain a robust estimator for $\sigma_{\rm z,0}$, we limited the fit range to [0,R$_{\rm 90}$] for the full galaxies. 
For the GMM discs we additionally excluded the 2~kpc innermost region since the decomposition leads to central dips.  

The right column of Fig.~\ref{fig_faceon_and_edgeon_radialparam} gives v$_{\rm los}$ as a function of the position along the major axis, x. 
The colors quantify the position along the minor axis, y. From these figures, we derived v$_{\rm max}$ as the maximum of $|$v$_{\rm los}$$|$. 
The disc and full galaxy show a typical rotation curve for a late type galaxy, 
with velocities decreasing with increasing distance from the equatorial plane, 
for a fixed position along the major axis. 

Table~\ref{table2} gives all the quantities we have analysed in this study for the eighteen 
galaxies and their GMM defined discs and spheroids. 

\begin{table*}
\centering
\begin{tabular}{ccccccccccccc}
\hline
Sim & j & b/a & c/a & $\epsilon_{\rm ms}$ & $\lambda$ & R$_{\rm 50}$ & n & time$_{\rm 50}$ & [Fe/H]$_{\rm 50}$ & [O/Fe]$_{\rm 50}$ & $\sigma_{\rm z,0}$ & z$_{\rm 0}$\\
 & kpc km s$^{\rm -1}$ & & & & & kpc & & Gyrs & dex & dex & km $s^{\rm -1}$ & kpc\\
 \hline
g9.59e10 & 77 & 0.96 & 0.56 & 0.42 & 0.33 & 4.78 & 1.24 & 8.78 & -1.40 & 0.05 & 26.01 & 1.07\\ 
g1.05e11 & 77 & 0.61 & 0.40 & 0.38 & 0.34 & 4.68 & 1.05 & 9.26 & -1.22 & 0.05 & 30.64 & 0.89\\
g1.37e11 & 77 & 0.86 & 0.46 & 0.52 & 0.35 & 3.19 & 0.89 & 4.85 & -0.70 & 0.13 & 39.88 & 0.76\\
g2.42e11 & 260 & 0.98 & 0.39 & 0.61 & 0.58 & 4.24 & 0.83 & 6.58 & -0.50 & 0.13 & 47.93 & 0.72\\
g3.59e11 & 127 & 0.70 & 0.48 & 0.33 & 0.29 & 4.48 & 1.49 & 10.25 & -0.73 & 0.15 & 50.17 & 1.36\\
g3.71e11 & 152 & 0.78 & 0.48 & 0.50 & 0.49 & 2.07 & 6.42 & 8.90 & -0.21 & 0.16 & 84.49 & 0.76\\
g4.90e11 & 361 & 0.79 & 0.30 & 0.68 & 0.62 & 6.37 & 0.87 & 8.14 & -0.77 & 0.14 & 45.27 & 0.91\\
g5.02e11 & 284 & 0.98 & 0.45 & 0.52 & 0.53 & 3.49 & 1.58 & 7.27 & -0.33 & 0.22 & 65.45 & 0.77\\
g5.31e11 & 151 & 0.98 & 0.41 & 0.44 & 0.36 & 2.45 & 1.82 & 8.44 & -0.21 & 0.16 & 79.80 & 0.73\\
g5.36e11 & 162 & 0.88 & 0.70 & -0.02 & 0.35 & 3.91 & 1.53 & 11.55 & -0.53 & 0.16 & 74.27 & 1.38\\
g5.38e11 & 371 & 0.97 & 0.38 & 0.69 & 0.65 & 3.14 & 1.89 & 6.64 & -0.28 & 0.20 & 78.76 & 0.70\\
g5.46e11 & 255 & 0.89 & 0.33 & 0.66 & 0.53 & 5.21 & 1.06 & 8.13 & -0.69 & 0.14 & 43.69 & 0.82\\
g6.96e11 & 290 & 0.96 & 0.60 & 0.38 & 0.41 & 4.12 & 1.46 & 11.23 & -0.23 & 0.19 & 89.48 & 0.84\\
g7.08e11 & 515 & 0.95 & 0.43 & 0.53 & 0.66 & 3.19 & 2.24 & 7.01 & -0.04 & 0.15 & 102.77 & 0.78\\
g7.55e11 & 453 & 0.94 & 0.37 & 0.76 & 0.63 & 3.01 & 2.75 & 7.15 & -0.19 & 0.21 & 85.91 & 0.74\\
g7.66e11 & 220 & 0.94 & 0.65 & 0.31 & 0.40 & 1.74 & 2.28 & 8.14 & -0.04 & 0.22 & 171.08 & 0.59\\
g8.13e11 & 307 & 0.94 & 0.53 & 0.46 & 0.63 & 1.36 & 1.67 & 4.79 & 0.11 & 0.20 & 205.78 & 0.47\\
g8.26e11 & 573 & 0.95 & 0.36 & 0.77 & 0.75 & 2.30 & 3.28 & 4.41 & -0.05 & 0.21 & 127.99 & 0.60\\
\hline
g9.59e10 & 295 & 0.61 & 0.26 & 0.64 & 0.73 & 8.92 & 0.62 & 9.83 & -1.38 & 0.05 & 23.38 & 0.73\\
g1.05e11 & 200 & 0.68 & 0.36 & 0.46 & 0.62 & 6.24 & 0.96 & 9.06 & -1.23 & 0.06 & 28.90 & 0.82\\
g1.37e11 & 231 & 0.90 & 0.34 & 0.70 & 0.68 & 4.69 & 0.83 & 6.16 & -0.60 & 0.09 & 28.78 & 0.64\\
g2.42e11 & 475 & 0.97 & 0.29 & 0.71 & 0.84 & 5.12 & 0.50 & 8.13 & -0.35 & 0.12 & 35.25 & 0.65\\
g3.59e11 & 552 & 0.71 & 0.38 & 0.47 & 0.66 & 10.85 & 1.10 & 8.97 & -1.00 & 0.14 & 52.64 & 1.29\\
g3.71e11 & 1586 & 0.60 & 0.22 & 0.78 & 0.97 & 21.12 & 0.26 & 8.76 & -0.87 & 0.13 & 39.00 & 1.45\\
g4.90e11 & 689 & 0.83 & 0.26 & 0.72 & 0.80 & 9.00 & 0.75 & 9.97 & -0.72 & 0.12 & 31.87 & 0.86\\
g5.02e11 & 650 & 0.96 & 0.22 & 0.72 & 0.89 & 4.66 & 1.00 & 9.63 & -0.08 & 0.13 & 40.39 & 0.68\\
g5.31e11 & 457 & 0.98 & 0.36 & 0.49 & 0.71 & 4.23 & 1.38 & 8.22 & -0.27 & 0.17 & 67.93 & 0.92\\
g5.36e11 & 370 & 0.95 & 0.57 & 0.28 & 0.73 & 3.60 & 0.96 & 12.30 & -0.49 & 0.17 & 76.47 & 1.49\\
g5.38e11 & 722 & 0.97 & 0.23 & 0.77 & 0.89 & 4.74 & 1.07 & 8.86 & -0.07 & 0.13 & 44.53 & 0.63\\
g5.46e11 & 637 & 0.93 & 0.25 & 0.74 & 0.80 & 8.84 & 0.84 & 10.12 & -0.64 & 0.12 & 29.61 & 0.78\\
g6.96e11 & 598 & 0.94 & 0.37 & 0.59 & 0.81 & 4.00 & 0.90 & 12.34 & -0.11 & 0.19 & 73.56 & 0.75\\
g7.08e11 & 502 & 0.96 & 0.42 & 0.54 & 0.67 & 3.33 & 2.17 & 6.59 & -0.07 & 0.17 & 102.56 & 0.78\\
g7.55e11 & 968 & 0.93 & 0.26 & 0.73 & 0.86 & 5.48 & 1.89 & 9.04 & -0.07 & 0.14 & 50.85 & 0.73\\
g7.66e11 & 596 & 0.93 & 0.23 & 0.70 & 0.93 & 1.97 & 0.99 & 10.53 & 0.34 & 0.12 & 75.98 & 0.37\\
g8.13e11 & 785 & 0.93 & 0.20 & 0.78 & 0.91 & 2.27 & 1.29 & 6.97 & 0.33 & 0.10 & 76.49 & 0.36\\
g8.26e11 & 968 & 0.96 & 0.31 & 0.69 & 0.86 & 3.75 & 2.63 & 5.59 & 0.04 & 0.15 & 67.13 & 0.64\\
\hline
g9.59e10 & 30 & 0.96 & 0.64 & 0.37 & 0.18 & 3.91 & 0.75 & 8.61 & -1.41 & 0.05 & - & 1.16\\
g1.05e11 & 9 & 0.51 & 0.39 & 0.24 & 0.19 & 3.75 & 0.85 & 9.28 & -1.22 & 0.05 & - & 0.95\\
g1.37e11 & 18 & 0.82 & 0.54 & 1.09 & 0.15 & 2.75 & 0.75 & 4.53 & -0.74 & 0.14 & - & 0.81\\
g2.42e11 & 19 & 0.99 & 0.55 & 0.46 & 0.10 & 3.20 & 1.46 & 4.30 & -0.73 & 0.19 & - & 0.82\\
g3.59e11 & 39 & 0.70 & 0.52 & 0.25 & 0.16 & 3.69 & 0.80 & 10.49 & -0.71 & 0.15 & - & 1.37\\
g3.71e11 & 91 & 0.94 & 0.58 & 0.48 & 0.29 & 1.70 & 1.42 & 8.91 & -0.20 & 0.16 & - & 0.74\\
g4.90e11 & 74 & 0.71 & 0.37 & 0.57 & 0.26 & 4.40 & 0.65 & 7.08 & -0.83 & 0.15 & - & 0.95\\
g5.02e11 & 26 & 0.97 & 0.64 & 0.36 & 0.09 & 2.76 & 1.95 & 6.38 & -0.57 & 0.25 & - & 0.84\\
g5.31e11 & 47 & 0.99 & 0.50 & 0.37 & 0.23 & 1.80 & 1.06 & 8.61 & -0.17 & 0.15 & - & 0.63\\
g5.36e11 & 139 & 0.77 & 0.65 & -0.33 & 0.29 & 4.47 & 2.40 & 10.37 & -0.68 & 0.15 & - & 1.22\\
g5.38e11 & 40 & 0.98 & 0.64 & 0.36 & 0.14 & 2.09 & 2.06 & 5.25 & -0.49 & 0.24 & - & 0.78\\
g5.46e11 & 78 & 0.85 & 0.40 & 0.55 & 0.24 & 3.88 & 0.91 & 7.43 & -0.73 & 0.15 & - & 0.84\\
g6.96e11 & 32 & 0.97 & 0.74 & 0.20 & 0.18 & 4.27 & 2.70 & 9.94 & -0.37 & 0.19 & - & 0.97\\
g7.08e11 & 406 & 0.95 & 0.47 & 0.52 & 0.62 & 2.89 & 2.37 & 4.79 & -0.32 & 0.27 & - & 0.80\\
g7.55e11 & 36 & 0.98 & 0.62 & 0.39 & 0.12 & 1.88 & 2.12 & 6.23 & -0.31 & 0.23 & - & 0.75\\
g7.66e11 & 59 & 0.95 & 0.71 & 0.25 & 0.13 & 1.56 & 3.38 & 7.49 & -0.24 & 0.24 & - & 0.68\\
g8.13e11 & 113 & 0.94 & 0.77 & 0.24 & 0.37 & 1.09 & 1.34 & 4.02 & -0.04 & 0.26 & - & 0.51\\
g8.26e11 & 76 & 0.93 & 0.71 & 0.50 & 0.30 & 1.34 & 1.10 & 3.27 & -0.27 & 0.29 & - & 0.55\\
\hline
\end{tabular}
\caption{Specific angular momentum (j), median-to-major semiaxes ratio (b/a), minor-to-major semiaxes ratio (c/a), 
\emph{measured} ellipticity ($\epsilon_{\rm ms}$), spin ($\lambda$), face-on half mass radius (R$_{\rm 50}$), S\'{e}rsic index (n), 
half mass formation time (time$_{\rm 50}$), half mass metallicity ([Fe/H]$_{\rm 50}$), half mass $\alpha$-enhancement ([O/Fe]$_{\rm 50}$)
central vertical velocity dispersion ($\sigma_{\rm z,0}$) and scale height (z$_{\rm 0}$).
The top block corresponds to the full galaxies, the central one to the GMM discs, and the bottom to the GMM spheroids.}
\label{table2}
\end{table*}

\end{document}